\documentclass[journal,twoside,web]{ieeecolor}
\usepackage{generic}
\usepackage{cite}
\usepackage{amsmath,amssymb,amsfonts}
\usepackage{graphicx}
\usepackage{algorithm,algorithmic}
\usepackage{hyperref}
\hypersetup{hidelinks=true}
\usepackage{bm}
\usepackage{subfigure}
\usepackage{subcaption}
\newtheorem{Definition}{Definition}

\newtheorem{Proposition}{Proposition}

\newtheorem{Theorem}{Theorem}
\newtheorem{Lemma}{Lemma}
\newtheorem{remark}{Remark}
\newtheorem{example}{Example}
\newtheorem{Corollary}{Corollary}
\usepackage{textcomp}
\usepackage{caption}

\DeclareCaptionFormat{blueformat}{\textcolor{black}{#1#2#3}}

\usepackage{etoolbox}
\AtBeginEnvironment{algorithm}{\color{black}}
\def\BibTeX{{\rm B\kern-.05em{\sc i\kern-.025em b}\kern-.08em
    T\kern-.1667em\lower.7ex\hbox{E}\kern-.125emX}}
\begin{document}
\title{Privacy Guarantee for Nash Equilibrium Computation of Aggregative Games Based on Pointwise Maximal Leakage
}
\author{
Zhaoyang Cheng,~~Guanpu Chen, \IEEEmembership{Member,~IEEE},~~Tobias J. Oechtering, \IEEEmembership{Senior Member,~IEEE}, \\and Mikael Skoglund,  \IEEEmembership{Fellow,~IEEE}
\thanks{Zhaoyang Cheng, Tobias J. Oechtering, and Mikael Skoglund are with the School of Electrical Engineering and Computer Science, KTH Royal Institute of Technology, Stockholm 100 44, Sweden.
        {\tt\small zhcheng@kth.se, oech@kth.se, skoglund@kth.se}
        }
\thanks{Guanpu Chen is with the School of Automation, Southeast University, Nanjing 210096, China.
        {\tt\small guanpu\_chen@seu.edu.cn}
        }
}

\maketitle
\vspace{-5pt}
\begin{abstract}
Privacy preservation has served as a key metric in designing Nash equilibrium (NE) computation algorithms. Although differential privacy (DP) has been widely employed for privacy guarantees, it does not exploit prior distributional knowledge of datasets and is ineffective in assessing information leakage for correlated datasets. To address these concerns, we establish a pointwise maximal leakage (PML) framework when computing NE in aggregative games. By incorporating prior knowledge of players' cost function datasets, we obtain a precise and computable upper bound of privacy leakage with PML guarantees. {\color{black}In the entire view, we show PML refines DP by offering a tighter privacy guarantee, enabling flexibility in designing NE computation with prior knowledge.} Also, in the individual view, we reveal that the lower bound of PML can exceed the upper bound of DP by constructing specific correlated datasets. The results emphasize that PML is a more proper privacy measure than DP since the latter fails to adequately capture privacy leakage in correlated datasets. Moreover, we conduct experiments with adversaries who attempt to infer players' private information to illustrate the effectiveness.

\end{abstract}

\vspace{-10pt}
\section{Introduction}












The computation of Nash equilibria (NE) has attracted significant attention in recent years due to its wide-ranging applications such as smart grids, economics, and traffic systems  \cite{gadjov2018passivity,lu2020online,chen2021distributed,deng2024distributed,chen2024approaching}. The aggregative game is a representative model in these contexts, wherein each player's cost function is influenced by an aggregate term reflecting the collective information of all players' strategies \cite{koshal2016distributed,liang2017distributed,paccagnan2018nash,belgioioso2020distributed,carnevale2024tracking}. The aggregate term is often inaccessible to the individual, necessitating players' estimation through the NE-computing process. To facilitate this, players iteratively share information with each other, and the computation of NE in aggregative games fundamentally relies on information exchanges among players.

Notably, privacy risk is exposed by information exchanges, which may inadvertently lead to the disclosure of private information. The privacy risk has been effectively explored in many practical situations, such as system identification and social welfare problems with novel privacy-preserving mechanisms \cite{tan2023Cooperative,yan2025social}. Addressing these privacy challenges is therefore critical for ensuring the security and applicability of {\color{black}aggregative games in practical computation \cite{ye2023distributed}. }{\color{black}Differential privacy (DP) \cite{dwork2006differential,dwork2010differential,dwork2014algorithmic} has been applied to algorithm design as a privacy guarantee, due to its explicit mathematical framework and easy applicability. 
Recent studies have integrated DP into game models \cite{ye2021differentially,wang2022differentially,wang2024differentially,wang2024ensuring,guo2025differentially}.
Recently, NE computation for aggregative games has been widely developed to maintain bounded privacy leakage while achieving high convergence accuracy in the presence of both Laplace noise \cite{ye2021differentially} and Gaussian noise \cite{wang2022differentially}.} Besides, with the bounded DP guarantee, exact convergence in NE computation has been further achieved through a novel aggregate estimation process \cite{wang2024differentially}. The line of research has also been extended to involve directed communication graphs \cite{wang2024ensuring,guo2025differentially}.

{\color{black}Nevertheless, DP has faced criticism and has been considered potentially vulnerable. A primary concern is that DP 
does not adequately assess information leakage in correlated datasets, as first pointed out in \cite{kifer2011no}.} For instance, in medical databases where sensitive attributes of different individuals may be interdependent,  the private health information may be inferred by attackers even if DP mechanisms are deployed \cite{saeidian2024evaluating}. In social networks where users’ preferences are correlated due to familial or social ties, a DP mechanism may fail to preserve private information from one node  \cite{kayes2017privacy}. {\color{black}Another significant factor is that DP does not exploit prior distribution knowledge if available, which may result in a poor privacy guarantee \cite{kifer2011no}. Even if the distributions of datasets are independent, the real information leakage of datasets with different priors
may vary under the same algorithm \cite{jiang2020local}. Consequently, inherent correlations may significantly exacerbate privacy leakage risks, and available prior information may influence the precision of the privacy guarantee, where DP exhibits shortcomings in these aspects.}

Pointwise Maximal Leakage (PML) has recently emerged as a promising approach, rooted in quantitative information flow \cite{saeidian2023pointwise}.
{\color{black}As its name suggests, PML generalizes the concept of maximal leakage \cite{issa2019operational} from a pointwise perspective of the output data. Specifically, PML quantifies information leakage from datasets with prior distributions to each possible observation output by the worst-case ratio between two scenarios—whether the output is observed by an adversary or not. This intrinsic link to adversarial detection provides a precise explanation of the privacy guarantee. 
The characteristic with adversary models 
underscores the operational meaning of PML, making it effective with prior distributions or correlated datasets \cite{saeidian2023rethinking,Grosse2024Extremal}. 
Moreover,  an equivalent formulation of PML \cite{saeidian2023pointwisegeneral} presents a potentially efficient and practical approach for assessing privacy guarantees.}

{\color{black}
Although the DP was widely adopted as the privacy guarantee in current privacy-preserving NE computation, it suffers from limitations when datasets are correlated or prior distributions are available. To this end, we develop a PML framework for privacy guarantees when computing NE in aggregative games, and conduct a comparative analysis between PML and DP.} Our contributions are summarized as follows:

\begin{itemize} \item  We propose a PML framework for assessing the privacy guarantee of NE computation algorithms in aggregative games. {\color{black}When the prior distribution is available, we derive a precise and computable upper bound of privacy leakage with PML guarantees for NE computation algorithms (Theorem \ref{th::determinsistic game upper bound}). Compared with the previous works focusing on one-step mappings \cite{saeidian2023pointwise,saeidian2023pointwisegeneral,saeidian2023rethinking}, we develop the PML to iterative algorithms in aggregative games.}

\item  In the entire view,  we show that the privacy guarantee with PML is tighter than that with DP,  given the prior distribution of players' datasets (Theorem \ref{th::compare_DP_PML_upper_bound}). {\color{black}We claim that PML refines DP in the entire view for NE computation with available prior distribution. It enables flexibility in designing NE computation algorithms for better tradeoff among privacy, accuracy,  and other considerations.}

\item  In the individual view, we construct specific correlated datasets of players' cost functions such that the lower bound guaranteed by PML exceeds the upper bound guaranteed by DP (Theorem \ref{th::compare_DP_PML_lower_boundd}). {\color{black} This reveals that PML remains a proper privacy guarantee, while DP underestimates the actual privacy leakage in correlated settings. By advancing the comparison from one-step mechanisms to iterative algorithms and aggregative games, our analysis makes PML applicable and proper for privacy assessment.}

\item We conduct experiments in contagious disease problems by employing a privacy-preserving NE computation algorithm. {\color{black}We investigate concrete adversaries attempting to infer private information through gradient matching. By applying our PML guarantees, we demonstrate that PML provides a more accurate assessment of privacy risks than the assessments by traditional DP in games with correlated datasets or available prior distribution. Additionally, the PML guarantees enable algorithm designers for better tradeoffs between privacy and accuracy.}

\end{itemize}

The organization of the paper is: Section II shows the problem formulation for the game model, privacy definitions for PML and DP.  Section III proposes a PML framework for privacy analysis when computing NE in aggregative games. Section IV compares the bounds of PML guarantees and DP guarantees in both the entire view and the individual view.  Section V gives numerical experiments to evaluate our theoretical results, followed by the conclusions in Section VI.

\textit{Notations:} We adopt the notational conventions: $\mathbb{R}_{+}=[0,\infty)$ and $[N]=\{1, \ldots, N\}$. Let $\mathbb{R}^d$ denote the Euclidean space of dimension $d$. Take $1_d$ as the $d$-dimensional column vector with all entries equal to $1$, and $I_d$ as the identity matrix of dimension $d$. 
Let $\|x\|$ be the standard Euclidean norm of a vector $x$, and $\|x\|_1$ represent the $L_1$ norm of a vector $x$. 
Let $A^T$ denote the transpose of a matrix $A$. For two vectors $u$ and $v$ with the same dimension, we use $u \leq v$ to represent the relationship that every entry of the vector $u$ is no greater than the corresponding entry of $v$. For a random variable $X$ with distribution $P_X$, take $\operatorname{supp}\left(P_X\right)\triangleq\left\{x \in \mathcal{X}: P_X(x)>0\right\}$ as as the support sets of $X$. 
{\color{black}Let $X$ be a topological space with topology $\tau$. The Borel $\sigma$-algebra on $X$, denoted by $\mathcal{B}(X)$, is defined as the $\sigma$-algebra generated by all open sets in $X$, that is,
$\mathcal{B}(X) := \sigma(\tau).$
A measurable space $(E,\mathcal{E})$ is called a standard Borel space if there exists a Polish space $X$ such that $(E,\mathcal{E})$ is measurably isomorphic to $(X,\mathcal{B}(X))$, where $\mathcal{B}(X)$ denotes the Borel $\sigma$-algebra on $X$.}

\vspace{-5pt}
\section{Problem Formulation and Privacy Definitions}
{\color{black}In this section, we first formulate the aggregative game model and describe the NE computation algorithm. We then present the definitions of the PML and the DP.}
\subsection{Game Model and NE Computation}


Consider an aggregative game with  $ N $ players, where players are indexed by $[N]=\{1, \ldots, N\}$. For each $i\in[N]$,  player $i$ takes a strategy $x_i$ in a local constraint set $\Omega_i \subseteq \mathbb{R}^n$. Denote $x\triangleq\left(x_1^T, \ldots, x_N^T\right)^T$ as the strategy profile for all players,  $ x_{-i} \triangleq\left(x_1^T, \ldots, x_{i-1}^T , x_{i+1}^T , \ldots , x_N^T\right)^T  $ as the strategy profile for all players except player $ i $, and $ \Omega\triangleq \prod_{i=1}^N \Omega_i \subseteq \mathbb{R}^{Nn}$ as all players' constraint set. 
Player $i$ has its own cost function \begin{equation}
\label{eq::costfunction}
   f_{i}(\cdot): \mathbb{R}^{Nn}\rightarrow\mathbb{R},
\end{equation}
where $f_{i}(x_i,x_{-i})=\bar{f}_i\left(x_i,\delta( x)\right)$ and  $\delta(x) \triangleq \frac{1}{N} \sum\limits_{i=1}^N x_i$ is the aggregate term of all players. {\color{black}$\bar{f}_i: \mathbb{R}^n\times\mathbb{R}^n\rightarrow\mathbb{R}$ formalizes the cost function in aggregative games, which is
influenced by the player's own strategy and the aggregate term.} Take $f=\{f_1,\dots, f_N\}$ as the set of all players' cost functions, and  $f_{-i}=\{f_1,\dots,f_{i-1},f_{i+1},\dots, f_N\}$ as the cost function set except player $i$. 
Denote the pseudo-gradient of the $i$th player as $F_i(x_i,\delta(x))\triangleq \nabla_{x_i} \bar{f}_{i}(x_i,\delta( x)). $
Given  $x_{-i}$, the objective of the $i$th player is to solve the following optimization problem:
\begin{equation}\label{problemformulation}
\begin{aligned}
\min_{x_i\in\Omega_i} f_i\left(x_i,x_{-i}\right).
\end{aligned}
\end{equation}
On this basis, the aggregative game is defined as a triple $G=\{[N],\Omega,  f\}$.
NE \cite{wang2024differentially,ye2021differentially} is a core concept of the game model, defined as follows.
\begin{Definition}[Nash equilibrium]
	 A strategy profile $x^*=\left(x_i^*, x_{-i}^*\right) \in \Omega$ is an NE of $G$ if for all $i \in [N]$, $ x_i \in \Omega_i $,
	$$
	\begin{aligned}
	&  f_i\left(x_i^*, x_{-i}^*\right) \leq f_i\left(x_i, x_{-i}^*\right).\\
	\end{aligned}
	$$
\end{Definition}
NE is a strategy profile where no player can reduce its cost by unilaterally changing its own strategy.

Consider a series of classic NE-computing algorithms within noise-adding mechanisms to protect players' private information, $f$, in aggregative games \cite{wang2024differentially,ye2021differentially,wang2022differentially}. Each player has no direct access to others' cost functions or the aggregate term. Then each player needs to construct a local estimate of the aggregate term. Each player adds noise to the estimate term and broadcasts the obscured estimate to others, which is also called observation. For convenience, take $x_i^k$, $v_i^k$, and $o_i^k$ as the strategy,  the estimate of the aggregate term,  and the observation item of the $i$th player in iteration $k$, respectively. Let $\bm x=\{x^k\}_{k=0}^T$, $\bm v=\{v^k\}_{k=0}^T$, and $\bm o=\{o^k\}_{k=0}^T$ denote the strategy sequence, the estimated sequence, and the observation sequence from iteration $0$ to $T$, where  $x^k=\{x_i^k\}_{i=1}^N$, $v^k=\{v_i^k\}_{i=1}^N$, and $o^k=\{o_i^k\}_{i=1}^N$, respectively. 

Specifically, in the aggregative game at iteration $k$, the $i$th player only knows its own cost function $f_i$, strategy $x_i^k$, estimated term $v_i^k$, and observation  $o^k$. First, the $i$th player adds noise $\zeta_i^k$ to its own estimated aggregated term $v_i^k$, and the observation becomes $o_i^k=v_i^k+\zeta_i^k$. {\color{black}Second, the $i$th player updates its strategy $x_i^{k+1}$  by $x_i^{k+1}=h_x(x_i^k,v_i^k,\lambda^k,f_i)$, where $h_x: \mathbb{R}^n \times \mathbb{R}^n \times \mathbb{R} \times E_f \rightarrow \mathbb{R}^n$ and $\lambda^k$ is the stepsize. $h_x$ and $h_v$ in Algorithm~$\mathcal{A}$ are assumed to be measurable. Further, in order to get a correct estimation of the aggregated term $\delta(x)$, the $i$th player updates its estimation by $v_i^{k+1}=h_v(x_i^{k+1},x_i^k,v_i^k,o^k,\gamma^k)$, where $h_v:\mathbb{R}^n\times\mathbb{R}^n\times\mathbb{R}^n\times\mathbb{R}^n\times\mathbb{R}\rightarrow\mathbb{R}^n$ and $\gamma^k$ is the stepsize.}  We initialize $x_i^0 \in \Omega_i$ and $v_i^0=x_i^0$ for $i\in[N]$. The iterative algorithm for each player $i
\in[N]$ is summarized as algorithm $\mathcal{A}$ in the following:

\begin{algorithm}[ht]
\caption{\color{black}Privacy-Preserving NE Computation Algorithm}
\begin{algorithmic}[1]\color{black}
\REQUIRE For player $i\in [N]$, $x_i^0 \in \Omega_i, v_i^0 = x_i^0$. Take stepsizes $\lambda^k>0, \gamma^k>0$, and max iterations $T$.\\

\ENSURE For $k = 0, 1, \dots, T$, player $i$ does:

\STATE Generate the noise $\zeta_i^k$, and broadcast
$o_i^k = v_i^k + \zeta_i^k$.
\STATE Update its strategy:
$x_i^{k+1}=h_x(x_i^k,v_i^k,\lambda^k,f_i).$
\STATE Update the aggregate estimate:
$$v_i^{k+1}=h_v(v_i^k,x_i^{k+1},x_i^k,o^k,\gamma^k).$$
\end{algorithmic}
\end{algorithm}

In algorithm $\mathcal{A}$, the observation  $o_i^k$ obtained by the noise-adding mechanism is specifically designed to preserve privacy. Meanwhile, the strategy update function $h_x$  and the estimation aggregate term update function $h_v$  are structured to compute the NE in aggregative games. 
{\color{black}Note that the strategy update function $h_x$ could be based on the gradient play, the best response dynamics, and the fictitious play, such as  $x_i^{k+1}=x_i^k-\lambda^k\nabla_{x_i}\bar{f}_i(x_i^k,v_i^k)$ and $
x_i^{k+1}=\arg\min_{x_i\in\Omega_i}f_i(x_i,v_i^k)
$ \cite{ye2023distributed}. }
However, these NE computation requirements and the privacy requirements may exhibit an inherent conflict: while adding noise mitigates information leakage, it simultaneously risks compromising data transmission accuracy and undermining the algorithm's ability to achieve a desirable accuracy of games. Consequently, existing research efforts predominantly focus on resolving these tradeoffs among privacy, accuracy, regret, and so on \cite{ye2021differentially,wang2024differentially,lin2023statistical}.

  
\begin{remark}
Privacy-preserving distributed NE computation algorithm is one of the most widely analyzed iterative algorithms in aggregative games. There is a communication network among these players, and each player sends messages to its neighbors through the network.  Consider that information exchange is based on the communication weight matrix $\bm L=\left\{L_{i j}\right\}$, where $L_{i j}>0$ if players $j$ and $i$ can directly communicate with each other and $L_{i j}=0$ otherwise. Take $\mathcal{E}_i$ as the neighbor set of player $i\in [N]$, which is the collection of players $j$ such that $L_{ij}>0$.
{\color{black} In \cite{ye2021differentially}, the strategy update is $x_i^{k+1}=x_i^k-\lambda^k F_i(x_i^k,v_i^k)$, and the estimated aggregated term update is $v_i^{k+1}=\sum\limits_{j\in\mathcal{E}_i}L_{ij}o_j^k +x_i^{k+1}-x_i^k$,  where $\lambda^k=cq^k$. Besides, in \cite{wang2024differentially},  the strategy update is $x_i^{k+1}=  \text{Proj}_{\Omega_i}\left[x_i^k-\lambda^k F_i\left(x_i^k, v_i^k\right)\right]$, and the estimated aggregated term update is $v_i^{k+1}=  v_i^k+\gamma^k \sum\limits_{j \in \mathcal{E}_i} L_{i j}\left(o_j^k-o_i^k\right) +x_i^{k+1}-x_i^k$, where $\sum_{k=T}^{\infty} \gamma^k=\infty, \sum_{k=T}^{\infty} \lambda^k=\infty, \sum_{k=T}^{\infty}\left(\gamma^k\right)^2<\infty, \sum_{k=T}^{\infty}, $ and $\frac{\left(\lambda^k\right)^2}{\gamma^k}<\infty$.}
\end{remark}

{\color{black} 

\begin{remark}
\label{rem:aggregative_game_example}
Aggregative games constitute an important class of noncooperative games widely studied in economics, smart grids, and disease problems. For example, in economics \cite{deng2024distributed}, firm $i$ chooses production $x_i$, and the cost function takes the form
$\bar{f}_i(x_i,\delta(x))=c_i(x_i)-x_i^TP(\delta(x)),$
where the market price depends on the total supply represented by $\delta(x)$. 
In smart grid applications \cite{wang2024differentially}, each user's cost function is influenced by both its own energy consumption and the electricity prices, where electricity prices typically depend on the overall consumption level $\delta(x)$.
In vaccination games \cite{bauch2004vaccination}, each player selects a vaccination strategy $x_i$, and the cost depends on vaccination coverage,
$f_i(x_i,x_{-i})=A_ix_i+\delta(x)^TB_i(\mathbf{1}_n-x_i),$
where $\delta(x)$ represents the overall vaccination level.
\end{remark}

}



\subsection{Pointwise Maximal Leakage}
Recently, PML \cite{saeidian2023pointwise} has been proposed as a privacy measure inspired by quantitative information flow. {\color{black}As the name suggests, PML generalizes the concept of maximal leakage \cite{issa2019operational} in a pointwise view over each observation sequence $\bm o$ and quantifies the worst-case information leakage against outside adversaries.}  PML is obtained by assessing the risk posed by adversaries in two highly threatening views: the randomized functions view  \cite{issa2019operational} and the gain function view \cite{m2012measuring}. In the following, we introduce PML based on the gain function view, since it could be defined on the general database and is suitable for game theory with general cost function sets \cite{saeidian2023pointwisegeneral}.


Take $E_f$ and $E_o$ as the sets that contain all possible cost functions $f$ and observation sequences $\bm o$, respectively. Let  $\mathcal{F}$ and $\mathcal{O}$ be the Borel $\sigma$-algebra on $E_f$ and $E_o$ respectively. Suppose $(E_f,\mathcal{F})$ and $(E_o,\mathcal{O })$ are two standard Borel spaces. Take $F$ and $O$ as random variables taking values in $(E_f,\mathcal{F})$ and $(E_o,\mathcal{O})$ with distribution $P_F$ and $P_O$, respectively. Similarly, we can define $\mathcal{F}_i$ and $\mathcal{F}_{-i}$ generated by $f_i$ and $f_{-i}$, respectively. With this preparation, algorithm 
$\mathcal{A}$ could be viewed as a mapping $\mathcal{A}: E_f\to \mathcal{O}$.  We can further define a mapping $P_{O|F}:E_{f}\times \mathcal{O}\to [0,1]$ such that $P_{O|F=f}(\bm o)$ is the conditional probability of the observation sequence $\bm o\in\mathcal{O}$ given the cost function set $f\in E_f$. {Let $P_{FO}$ be the probability measure on the product space $(E_f\times  E_{o},\mathcal{F}\otimes\mathcal{O})$  with marginals $P_F$ and $P_{O}$.} Recalling that both $(E_o,\mathcal{O})$  and $(E_f,\mathcal{F})$ are standard Borel spaces, the joint distribution $P_{FO}$ can be disintegrated into the marginal $P_O$ and a transition probability kernel $P_{F|O}$ from $(E_o,\mathcal{O})$ to $(E_f,\mathcal{F})$.

Suppose that an adversary observes $\bm o$, and aims to guess the true $F$. Take $\hat{F}$ as the random variable of the adversary's guess with a distribution $P_{\hat{F}}$, and take $(E_{\hat{f}},\hat{\mathcal{F}})$ as the possible measurable space for $\hat{F}$. Analogously, the guessing process can be denoted by a mapping $P_{\hat{F}|O}:E_o\times\hat{\mathcal{F}}\to[0,1]$. Further, with any possible true cost function profile $f\in\mathcal{F}$ and any possible guess $\hat{f}\in\hat{\mathcal{F}}$, there should be an objective for the adversary to maximize, which could be described in the gain function view \cite{m2012measuring}. Specifically, consider $g: \mathcal{F}\times\hat{\mathcal{F}}\to \mathbb{R}^+$ as a possible gain function. Noticing adversary's preference is unknown, adversary's gain function may be constructed in different ways, for example, by the identity function or by other metrics and distances \cite{m2012measuring,saeidian2023pointwise}. 
{\color{black}\begin{remark}
The gain function $g$ models the adversary's inference objective in NE computation under privacy preservation. Different choices of $g$ correspond to different attack preferences. For example, if an adversary aims to exactly infer a player's private cost function, one may take 
$
g(f,\hat f)=\sum_{i=1}^N\mathbf 1\{f_i=\hat f_i\},
$
while if the adversary seeks to estimate a private parameter $\theta(f)$ as $\hat\theta(\hat{f})$, one may choose a gain based on estimation accuracy, e.g.,
$
g(f,\hat f)=-\|\theta(f)-\hat\theta(\hat{f})\|^2.
$
\end{remark}

}

Recall that $F$ and $\hat{F}$ are random variables according to $P_F$ and $P_{\hat{F}}$, respectively. This allows us to define random variables $g(F,\hat{f})$ and $g(F,\hat{F})$ induced by the above distributions. Specifically, take $\mathbb{E}[g(F,\hat{f})]$ as the adversary's expected gain regarding $F$. 
Given $\bm o$, let $\mathbb{E}[g(F,\hat{F})|O=\bm o]$ be the adversary's expected gain between $F$ and $\hat{F}$. Accordingly, the gain function $g$ should be picked from 
$\Gamma\triangleq\left\{ g: \mathcal{F}\times\hat{\mathcal{F}}\to \mathbb{R}^+|\sup\limits_{\hat{f}\in \hat{\mathcal{F}}}\mathbb{E}\left[g(F,\hat{f})\right]<\infty\right\}.$
Note that the requirement $\sup\limits_{\hat{f}\in \hat{\mathcal{F}}}\mathbb{E}\left[g(F,\hat{f})\right]<\infty$
implies that the value of expected gains can potentially be improved upon observing $\bm o$. More comprehensive discussions can be found in \cite{saeidian2023pointwisegeneral}. With the concept of expected gains, the worst-case for preserving privacy should be that the adversary maximizes its expected gain. Thus, 
take $\sup\limits_{\hat{f}\in\hat{\mathcal{F}}}\mathbb{E}[g(F,\hat{f})]$ as the worst-case value for preserving privacy when an adversary has no observation, and $\sup\limits_{P_{\hat{F}|O}} \mathbb{E}[g(F,\hat{F})|O=\bm o]$ as the worst-case value for preserving privacy when an adversary observes $\bm o$.

In order to measure the amount of information leaking with a prior distribution, we examine the worst-case ratio\footnote{We use the convention that $\frac{x}{0} =\infty$ if $x > 0$.} between two scenarios--whether the output sequence is observed or not by an adversary. Specifically, the PML of algorithm $\mathcal{A}$ from $F$ to $\bm o$ is defined as follows \cite{saeidian2023pointwise,saeidian2023pointwisegeneral}:

\begin{Definition}[Pointwise maximal leakage]\label{def::PML}
Given a prior $P_F$ and an observation sequence $\bm o$,  the PML of algorithm $\mathcal{A}$ from $F$ to $\bm o$ is
\begin{equation}\label{eq::PML}
    \begin{aligned}
\ell(F \rightarrow \bm o):
& =\log \sup\limits_{(E_{\hat{F}},\hat{\mathcal{F}}),g\in \Gamma} 
\frac{\sup\limits_{P_{\hat{F}|O}} \mathbb{E}[g(F,\hat{F})|O=\bm o]}{\sup\limits_{\hat{f}\in\hat{\mathcal{F}}}\mathbb{E}[g(F,\hat{f})]}.
\end{aligned}
\end{equation}
\end{Definition} 

Upon examining the formulation of (\ref{eq::PML}), we observe that PML is defined with the combination of the worst-case ratio between scenarios where the adversary observes $\bm o$ and where the adversary has no observation, and the suprema taken over all possible measurable guessing spaces $(E_{\hat{F}},\hat{\mathcal{F}})$ and all gain functions $g\in \Gamma$. First, it is clear that PML is intrinsically linked to the adversary's gain function, making it directly relevant to the specific task. This connection establishes PML as a framework for explainability. 
Second, the inclusion of the suprema ensures that the ratio captures the worst-case scenario and quantifies the largest amount of potential information leakage by considering a diverse array of potential adversaries. 
As a result, PML serves as a robust privacy measure, capable of addressing a wide spectrum of attack scenarios.  Furthermore, PML exhibits flexibility similar to DP, as it is a random variable that can be bounded and manipulated in various ways—for instance, by evaluating PML for an individual player’s dataset $F_i$.  This adaptability underscores PML’s operational meaning, making it applicable to a variety of practical scenarios \cite{saeidian2023pointwise}.

Obviously, the original definition of PML in equation (\ref{eq::PML}) appears complicated for direct application to practical analyses. Recent theoretical advancements reveal that PML admits an elegant equivalent characterization, by considering finite sets in \cite[Theorem 1]{saeidian2023pointwise} and general sets in \cite[Theorem 3]{saeidian2023pointwisegeneral}, respectively, which is shown in the following:

{\color{black}\begin{Proposition}[{\cite[Theorem 3]{saeidian2023pointwisegeneral}}]\label{Pro::1}
Given a prior $P_F$ and an observation sequence $\bm o$, the PML of  algorithm $\mathcal{A}$ from $F$ to $\bm o$ can be expressed as
   $\ell(F \rightarrow \bm o)=D_\infty(P_{F|O=\bm o}\|P_{F}),$
  where $D_\infty(P_{F|O=\bm o}\|P_{F})$
  denotes the R$\acute{e}$nyi divergence  of order infinity \cite{renyi1961measures,van2014renyi} of $P_{F|O=\bm o}$ from $P_{F}$.
    
\end{Proposition}
}

 {\color{black} According to the above proposition, PML has the following elegant formulation: $\ell(F \rightarrow \bm o)=\log \sup\limits_{f\in\operatorname{supp}\left(P_F\right)} \frac{P_{O|F=f}(\bm o)}{P_O(\bm o)}$.
 
Recall that the PML in Definition \ref{def::PML} is originally defined for each possible observation $\bm o$, which is emphasized by the pointwise view of maximal leakage, rather than an expectation of $\bm o$ \cite{issa2019operational}. It enables us to consider the almost-sure guarantee for PML. Specifically, noting $O$ is a random variable with the distribution $P_O$, the random variable $\ell(F\to O)$ is consequently well-defined by $O$ and $P_O$. To address the information leakage over $\ell(F\to O)$, distinct guarantees, such as the almost-sure guarantee, the tail-bound guarantee, and the average-case guarantee, can be applied. 
Among these, the almost-sure guarantee is the strictest one for assessing the information leakage with probability one. In this paper, we specifically investigate PML under the almost-sure guarantee framework \cite{saeidian2023pointwise,saeidian2023rethinking}.}

\begin{Definition}[$\epsilon$-pointwise maximal leakage]
Given  $\epsilon\geq0$ and a prior $P_F$, algorithm $\mathcal{A}$ is said to be $\epsilon$-PML, if 
 \begin{equation}\label{eq::epsilon_PML}
\mathbb{P}_{O\sim P_O}[\ell(F\to  O)\leq \epsilon] =1.
\end{equation}

\end{Definition}
\vspace{3pt}
{On this basis, algorithm $\mathcal{A}$ is $\epsilon$-PML if $\ell(F\to  O)\leq \epsilon$ holds almost surely for $O$. Recalling $\operatorname{supp}\left(P_O\right)=\left\{\bm o \in \mathcal{O}: P_O(\bm o)>0\right\}$, equation (\ref{eq::epsilon_PML}) can be rewritten as $\sup\limits_{\bm o\in\operatorname{supp}\left(P_O\right)} \ell(F\to \bm o) \leq \epsilon$.}
{\color{black}
 Specifically, PML also admits the composition property. 
 If independent mechanisms $\mathcal{M}_1, \dots, \mathcal{M}_T$ are $\epsilon_1$-, $\dots$, $\epsilon_T$-PML,  respectively, then the composed mechanism is $(\sum_{k=1}^T \epsilon_k)$-PML. 
}


\subsection{Differential Privacy}

DP serves as a popular privacy framework and has recently been widely applied to the algorithm design in optimization and game theory \cite{huang2015differentially,ye2021differentially,wang2022differentially,wang2024differentially,guo2025differentially}.
DP guarantees that an individual partaking in the privacy mechanism will not face substantially increased risk due to their participation. 
The expression of DP is based on adjacent datasets, namely, two datasets that only differ in one entry \cite{dwork2014algorithmic,dwork2010differential}. 
In the game model $G$, two function sets $f$ and $f'$ are adjacent if only one player's cost function is different \cite{ye2021differentially,wang2022differentially,wang2024differentially}, \textit{i.e.}, there exists $i\in [N] $ such that $f_i\neq f_i'$ but $f_j=f_j'$ for all $j\neq i$. The definition of DP regarding algorithm $\mathcal{A}$ is accordingly as follows \cite{wang2016relation}. 

\begin{Definition}[$\epsilon$-differential privacy]\label{def::DP}
Given $\epsilon\geq0$, algorithm $\mathcal{A}$ is said to be  $\epsilon$-DP if for
  any possible $\bm o\in\mathcal{O}$ and any adjacent datasets $f$ and $f'$, we have
\begin{equation*}
   \frac{P_{O|F=f}(\bm o)}{P_{O|F=f'}(\bm o)}\leq \exp (\epsilon).
\end{equation*}

\end{Definition}
\vspace{3pt}

The formulation of DP in Definition \ref{def::DP} is consistent with
 other common formulations. For example, 
according to the formulation in the field of optimization \cite{dwork2014algorithmic,ye2021differentially},  algorithm $\mathcal{A}$ is $\epsilon$-DP  if for any two adjacent datasets $f,f^{\prime}\in E_f$ and any observation sequence set $ \bm o_s\subseteq E_o$, 
\begin{align*}
    \mathbb{P}\left[\mathcal{A}\left(f\right) \in \bm o_s\right] \leq e^\epsilon\mathbb{P}\left[\mathcal{A}\left(f'\right) \in \bm o_s\right].
\end{align*}
All the above formulations illustrate the fact that, if the privacy guarantee $\epsilon $ is small enough,  then the likelihoods of all possible observation sequences $\bm o$ and all adjacent datasets are close, making it hard to distinguish between the two datasets with high probability. Noticing that an algorithm with $0$-DP  may result in meaningless outputs and significantly diminish its practicality,  we always suppose that $\epsilon\neq 0$.

Besides, DP has been extended from description for adjacent datasets to that for group datasets \cite{dwork2014algorithmic,dwork2006differential,soria2017individual}, in order to accommodate practical scenarios where datasets may differ in more than one entry. Conveniently, take $[N]_k=\left\{G|G\subseteq [N],|G|=k\right\}$ as all subsets of size $k$ in set $[N]$. Similar to the adjacent definition, two cost function sets $f$ and $f'$ differ at $k$ entries if only $k$ players' cost functions are different, i.e., there exists $G\in [N]_k $ such that $f_i\neq f_i'$  for all $ i\in G$ but $f_j\neq f_j'$  for all $j\in[N]/G$. The definition of $\epsilon$-DP for groups of size $k$ is shown in the following \cite{dwork2014algorithmic,dwork2006differential,soria2017individual}.

\begin{Definition}\label{def::DP for groups}[$\epsilon$-differential privacy for groups of size $k$]
Given $\epsilon\geq 0$, algorithm $\mathcal{A}$ is said to be $\epsilon$-DP for groups of size $k$, if for
  any possible $\bm o\in\mathcal{O}$ and any two datasets $f$ and $f'$ differ at $k$ entries, we have
\begin{equation*}
   \frac{P_{O|F=f}(\bm o)}{P_{O|F=f'}(\bm o)}\leq \exp (\epsilon).
\end{equation*}

\end{Definition}
\vspace{3pt}

Notice that in \cite{soria2017individual}, $\epsilon$-DP for groups is also called $\epsilon$-group DP. 
Upon observing the formulation of DP for groups in Definition \ref{def::DP for groups}, it broadens the privacy in the individual entry difference view to the group entry difference view. Obviously, $\epsilon$-DP for groups of size $1$ is equivalent to $\epsilon$-DP in Definition \ref{def::DP}, and  $\epsilon$-DP for groups of size $N$ describes {\color{black}scenarios} where two datasets may differ in all entries.
 
Further, the existing analysis has been given for the relation between DP for groups with DP. 
\begin{Proposition}[{\cite[Theorem 2.2]{dwork2014algorithmic}}]\label{pro::DP for groups}
 Any $\epsilon$-DP mechanism is also $k\epsilon$-DP for groups of size $k$. 
\end{Proposition}




{\color{black}Note correlation may naturally arise in aggregative games due to common environmental factors, social influences, or shared external signals. The relation among cost function datasets does not directly affect the DP guarantee in Definition \ref{def::DP}, and the privacy leakage under PML explicitly depends on the prior distribution, allowing correlations among players' datasets to be incorporated into the privacy analysis. Therefore, in the following sections, we develop a PML framework for privacy analysis in NE computation for aggregative games. }

\section{Privacy Guarantee with PML Framework}


In this section, we develop a  PML framework for the privacy analysis within the context of the classic NE-computing algorithm for aggregative games. 


Firstly, we provide a preliminary method to simplify the calculation of the bound with PML guarantee. The proof is provided in Appendix \ref{ap::le::PMLconvert}.


\begin{Lemma}\label{le::PMLconvert}
If $\inf\limits_{\bm o\in\operatorname{supp}\left(P_O\right)}\inf\limits_{f'\in\operatorname{supp}\left(P_F\right)}\mathbb{E}_{f\sim F}\left[\frac{P_{O|F=f}(\bm o)}{P_{O|F=f'}(\bm o)}\right]\geq \exp(-\epsilon),$ then
$\begin{aligned}
    \sup\limits_{\bm o\in\operatorname{supp}\left(P_O\right)}\exp \ell(F\to \bm o)\leq \exp(\epsilon).
\end{aligned}$
\end{Lemma}


In the following, we introduce some notation for the convenience of description. Consider a fixed realization of the cost function set $f$  and an observation sequence $\bm o$.
First, at each iteration $k$, define  \begin{equation}\label{eq::A(ofxvk)}
    \begin{aligned}
    A(\!o^{k},\!f| x^{k},\! v^{k}\!)\!\!=\!\!\{ x^{k\!+\!1}\!,\! v^{k\!+\!1}|&x_i^{k\!+\!1}\!=\!h_x(x_i^{k},v_i^{k},\lambda^{k},f_i),\\
    &v_i^{k\!+1}\!=\!h_v(v_i^k,x_i^{k\!+\!1},x_i^k,o^k,\gamma^k)\!\}
\end{aligned}
\end{equation}as
the set of possible $ x^{k+1}, v^{k+1}$ in algorithm $\mathcal{A}$   given $x^k$, $v^k$, $o^k$, and $f$. {\color{black}Besides, define \begin{equation}\label{eq::Aof}
    A(\bm o,\!f)\!=\! \left\{\!(\bm x,\bm v)|( x^{k}\!,\! v^{k})\!\in\! A(\! o^{k\!-\!1},\!f| x^{k\!-\!1},\! v^{k\!-\!1}\!)\!,\!k\geq 1\!\right\}
\end{equation} as the set of all possible strategy sequence $\bm x$ and estimated term $\bm v$ by given an observation sequence $\bm o$ and a cost function set $f$.} Notice that, for any given $\bm o,f, x^{k}$, and $ v^{k}$, $A(o^{k-1},f| x^{k-1}, v^{k-1})$ is a singleton set, and $A(\bm o,f)$ contains a single sequence with given initial value $ x^0$ and $ v^0$. Based on the above definition for a single $\bm o$ and $f$, it is straightforward to extend the definition of $A(\bm o,f)$ when $\bm o$ and $f$  are considered within the Borel $\sigma$-algebra $\mathcal{O}$ and $\mathcal{F}$, respectively. {\color{black}Second, let $Q_i^k(\bm o,\bm v,\bm v', M^k)=\exp(-\frac{||o_{i}^k-v_i^{k}||_1-||o_{i}^k-v_i^{\prime k}||_1}{M^k})$ be a difference value for the $i$th player at iteration $k$ among any possible $\bm o, \bm v$, $\bm v'$, and $M^k$. Further, take $O^k$ and $V^k$ as the random variables representing the observation $o^k$ and the estimate term $v^k$ at iteration $k$, respectively.} 

With the above notation, we can rewrite $\frac{P_{O|F=f}(\bm o)}{P_{O|F=f'}(\bm o)}$ with Laplace distributions in the following lemma,  proof of which can be found in Appendix \ref{ap::le::1}.

\begin{Lemma}\label{le::1}
    Suppose the noise $\zeta_i^k$ follows a Laplace distribution $\zeta_i^k\sim Lap(0,M^k)$. 
    For any possible cost function set $f,$ $f'$ and any possible observation sequence $\bm o $, 
    we have $
\frac{P_{O|F=f}(\bm o)}{P_{O|F=f'}(\bm o)}=\frac{\int_{(\bm x',\bm v')\in A(\bm o,f')}\Pi_{k=0}^T\Pi_{i=1}^NQ_i^k(\bm o,\bm v,\bm v', M^k)P_{O_i^{k}|F=f',V_i^k=v_i^{\prime k}}(o_i^k)d (\bm x',\bm v')}{\int_{(\bm x',\bm v')\in A(\bm o,f')}\Pi_{k=0}^T\Pi_{i=1}^NP_{O_i^{k}|F=f',V_i^{k}=v_i^{\prime k}}(o_i^k)d(\bm x',\bm v')}, $ 
where $(\bm x,\bm v)\in A(\bm o,f)$.
\end{Lemma}

Besides, {\color{black}at each iteration $k$, take $$\Delta(k,f,f',\bm o)=\sup\limits_{\scriptsize \substack{v^k\in A(\bm o,f) \\ v^{\prime k}\in A(\bm o,\!f')}}\|v^k-{v}^{\prime k}\|_1$$
 \vspace{-3pt}
 as the sensitivity of algorithm $\mathcal{A}$ with a fixed $f,f'$ and $\bm o$, which quantifies how distinguishable two game instances are at
iteration $k$.} Similarly, it is straightforward to extend the definition of $\Delta(k,f,f',\bm o)$ when $f$, $f'$, and $\bm o$  are considered in the Borel $\sigma$-algebra $\mathcal{F}$ and $\mathcal{O}$.  {\color{black} In aggregative games, it describes the maximum deviation between the aggregate term estimates $v^k$ with two different cost function profiles and the same observation sequence, which quantifies how distinguishable two game instances are at iteration $k$.}
Based on above formulation, we get a lower bound and an upper bound of $\frac{P_{O|F=f}(\bm o)}{P_{O|F=f'}(\bm o)}$ in the following lemma, whose proof can be found in Appendix \ref{ap::le::delta}.

\begin{Lemma}\label{le::delta}
Suppose  $\zeta_i^k\sim Lap(0,M^k)$. 
    For any possible $f,$ $f'$ and $\bm o $, we have 
    $$\!\exp\!\left(\!\!-\!\!\sum\limits_{k=0}^T\frac{\Delta\!(k,\!f,\!f'\!,\!\bm o)}{M^{k}}\!\!\right)\!\leq\!\frac{P_{O|F\!=\!f}(\!\bm o\!)}{P_{O|F\!=\!f'}(\!\bm o\!)}\!\leq \!\exp\!\left(\sum\limits_{k=0}^T\!\frac{\Delta\!(k,\!f,\!f'\!,\!\bm o)}{M^{k}}\!\!\right)\!\!.$$
    
\end{Lemma}
\vspace{3pt}

On this basis, we provide a PML framework to assess the privacy guarantee of the NE computation for aggregative games in the following theorem, whose proof is shown in Appendix \ref{ap::th::determinsistic game upper bound}.
\begin{Theorem}\label{th::determinsistic game upper bound}
    \!\!\! Suppose $\zeta_i^k\!\!\sim \!\! Lap(0,\!M^k)$. Given a prior $P_F$, if 
    \begin{equation}\label{eq::th::PML}
   \! \inf\limits_{\scriptsize \substack{\bm o\in\operatorname{supp}\left(\!P_O\!\right) \\ f'\in\operatorname{supp}\left(\!P_F\!\right)}}\!\mathbb{E}_{f\sim F}\!\left[\exp\!\left(\!-\!\sum\limits_{k=0}^T\frac{\Delta(k,f,f'\!,\!\bm o)}{M^k}\right)\!\right]\!\geq \!\exp(-\epsilon),       
    \end{equation}
    then algorithm $\mathcal{A}$ is $\epsilon$-PML.
\end{Theorem}

{
{\color{black}Theorem \ref{th::determinsistic game upper bound} shows a computable formulation for the privacy guarantee of NE-computing algorithms with any prior in the PML framework. The computed PML guarantee remains valid under independent or correlated datasets. }Given the specific prior $P_F$ for players' cost functions, we can provide the PML privacy guarantee through the infimum of $\mathbb{E}_{f\sim F}\!\left[\exp\!\left(\!-\!\sum\limits_{k=0}^T\frac{\Delta(k,f,f',\bm o)}{M^k}\right)\!\right]$. Although Theorem \ref{th::determinsistic game upper bound} is based on a finite number of iterations, $T$, it can be directly extended to the infinite case for $T\to \infty$, if the infimum of $\mathbb{E}_{f\sim F}\!\left[\exp\!\left(\!-\!\sum\limits_{k=0}^\infty\frac{\Delta(k,f,f',\bm o)}{M^k}\right)\!\right]$ is not $0$. {\color{black}The term $\frac{\Delta(k,f,f',\bm o)}{M^k}$ captures the normalized sensitivity at iteration $k$ and cost function profile $f$, and the designer needs to design the allocation over the prior $P_F$ and the time horizon $T$, which may lead to better privacy–accuracy tradeoff. 
Notably, this work represents a novel result to derive a precise and computable privacy guarantee within the PML framework, incorporating prior knowledge. Note that we are the first to establish the PML framework for privacy-preserving NE computing algorithms in aggregative games, while previous analyses of PML methods have predominantly focused on one-step mappings from datasets to observations \cite{saeidian2023pointwise,saeidian2023pointwisegeneral,saeidian2023rethinking}. } 

{\color{black} 
\begin{remark}
We characterize how to compute the corresponding PML guarantee for algorithm $\mathcal{A}$ with a finite iteration horizon. The next step should be how to negotiate the tradeoff between privacy and accuracy in computing NE. For example, the algorithm designer should have a privacy budget when computing NE, and one of the stopping conditions is that the computed privacy leakage exhausts the privacy budget.
\end{remark}
\begin{remark}
    Our result in Theorem 1 can also be interpreted as a compositional bound for the considered iterative algorithm. Specifically, if for each iteration $k$ we have 
$\inf_{\substack{\bm o\in\operatorname{supp}(P_O) \\ f'\in\operatorname{supp}(P_F)}} \mathbb{E}_{f\sim F}\left[ \exp\left( -\frac{\Delta(k,f,f',\bm o)}{M^k} \right) \right] \ge \exp(-\epsilon_k),$ then the composed mechanism satisfies $\inf_{\substack{\bm o\in\operatorname{supp}(P_O) \\ f'\in\operatorname{supp}(P_F)}} \mathbb{E}_{f\sim F}\left[ \exp\left( -\sum_{k=0}^T \frac{\Delta(k,f,f',\bm o)}{M^k} \right) \right] \ge \exp\left(-\sum_{k=0}^T \epsilon_k\right),$ which implies that the overall algorithm is $(\sum_{k=0}^T \epsilon_k)$-PML. This is consistent with the general composition property of PML, and one of our framework's advantages is that Theorem 1 often yields tighter bounds than directly summing $\epsilon_k$.
\end{remark}
}

Further, by Fatou's lemma, we can obtain a concise version by taking $\Delta'(k,f)=\sup_{\scriptsize \substack{\bm o\in\operatorname{supp}\left(\!P_O\!\right) \\ f'\in\operatorname{supp}\left(\!P_F\!\right)}}\Delta(k,f,f',\bm o)$  as follows:
\begin{Corollary}\label{co::1}
Suppose $\zeta_i^k\sim Lap(0,M^k)$. Given a prior $P_F$, if $$\mathbb{E}_{f\sim F}\left[\exp\left(-\sum\limits_{k=0}^T\frac{\Delta'(k,f)}{M^k}\right)\right]\geq \exp(-\epsilon),$$ then algorithm $\mathcal{A}$ is $\epsilon$-PML.
\end{Corollary}

{\color{black}
Moreover, if the supremum of $\sum\limits_{k=0}^T\frac{\Delta'(k,f)}{M^k}$ over $f$ is bounded, by taking $\Delta''(k)=\sup_{\scriptsize \substack{\bm o\in\operatorname{supp}\left(\!P_O\!\right) \\ f,f'\in\operatorname{supp}\left(\!P_F\!\right)}}\Delta(k,f,f',\bm o)$, we can get the following corollary.
\begin{Corollary}\label{co::2}
Suppose $\zeta_i^k\sim Lap(0,M^k)$. If $\sum\limits_{k=0}^T\frac{\Delta''(k)}{M^k}\leq \epsilon,$ then algorithm $\mathcal{A}$ is $\epsilon$-PML.
\end{Corollary}
}
{\color{black} \begin{example}
In many existing privacy analyses, the boundedness of the sum $\sum_{k=0}^T \frac{\Delta''(k)}{M^k}$ is important to be established. For instance, consider the algorithm in \cite{ye2021differentially} with strategy update $x_i^{k+1}=x_i^k-\lambda^k F_i(x_i^k,v_i^k)$ and aggregate estimate update $v_i^{k+1}=\sum_{j=1}^N\frac{1}{N}o_j^k +x_i^{k+1}-x_i^k$, where the privacy guarantee requires $\sum_{k=0}^T\frac{\Delta''(k)}{M^k}\leq \epsilon$. As shown in \cite[Theorem 3]{ye2021differentially}, if the pseudo-gradient is uniformly bounded, i.e., $|\nabla_{x_i} \bar{f}_{i}(x_i,\delta(x))|\leq C$, then one obtains
$\|v^k-{v}^{\prime k}\|_1\leq \|x^k-{x}^{k-1}\|_1+\|x^{\prime k}-{x}^{\prime k-1}\|_1\leq 2\lambda^{k-1}C$,
which implies $\Delta''(k) \leq 2\lambda^{k-1}C$. The problem then reduces to ensuring the boundedness of $\sum_{k=0}^T \frac{2\lambda^{k-1}C}{M^k}$ for $T\to \infty$. By selecting stepsizes and noise scales as $\lambda^k = c q^k$ and $M^k = d \bar{q}^k$ with $\bar{q} \in (q,1)$, the sum is bounded by $\frac{2cC\bar{q}}{d(\bar{q}-q)}$. Similar bounds can be derived for other aggregate estimate update rules \cite{wang2024differentially,wang2022differentially}.
\end{example}

\begin{remark}

    By taking the worst-case PML over all possible prior distributions, PML remains meaningful even when the algorithm designer does not obtain the true prior $P_F$. Specifically, Corollary 2 takes the supremum over all possible $P_F$, which removes the dependence on prior distribution and provides the same privacy guarantee as DP in the entire view. Thus, even if the true prior $P_F$ is not available, the worst-case PML guarantee over distributions can still serve as a useful privacy guarantee, just like DP in the entire view.
\end{remark}
   }

Both Corollary \ref{co::1} and Corollary \ref{co::2} provide simplified boundaries for the privacy guarantee. 
It is worth noting that Theorem \ref{th::determinsistic game upper bound} offers a tighter bound compared to these two corollaries. {\color{black}These privacy guarantees offer designers a range of options tailored to diverse preferences, balancing stricter privacy preservation and simplified implementation. Moreover, Corollary \ref{co::2} bears a resemblance to existing results in the DP framework in \cite[Lemma 2]{huang2015differentially}, regarding $f$ and $f'$  are not adjacent. Given that the DP may be inadequate in correlated datasets, it becomes crucial to undertake a detailed comparison of the privacy guarantees between the PML and the DP—a task that will be addressed in the subsequent section.}


{\color{black}

}

\section{Comparison between PML and DP}

In this section, we compare the performance of the PML framework and the DP framework. This analysis seeks to demonstrate that the PML framework not only outperforms the DP framework in general scenarios but also proves indispensable when dealing with correlated datasets.






\vspace{-10pt}
\subsection{In Entire View}



{\color{black}We focus on the information leakage from entire datasets, which leads to the comparison between PML and DP in the \textit{entire} view. }
Regarding the formulation of $\ell(F \rightarrow \bm{o})$ in Definition \ref{def::PML}, PML serves as a privacy measure that assesses the privacy leakage of the variable $F$. Specifically, it assesses leakage across the entire dataset under the prior $P_F$, rather than focusing on the differences between adjacent datasets. 
Additionally, as stated in Definition \ref{def::DP for groups}, DP for groups of size $N$ considers differences in the view of the entire dataset, whereas traditional DP, in Definition \ref{def::DP}, is concerned with variations between adjacent datasets. Consequently, it is both logical and meaningful to compare the privacy guarantees provided by PML with those offered by DP for groups of size $N$ from the entire view. To this end, we present the following theorem, with the proof available in Appendix \ref{ap::th::compare_DP_PML_upper_bound}.

\begin{Theorem}\label{th::compare_DP_PML_upper_bound}
  Consider any prior $P_F $ and privacy budget $\epsilon_1$.

   (1) If algorithm $\mathcal{A}$ is $\epsilon_1$-DP for groups of size $N$, then  algorithm $\mathcal{A}$ is $\epsilon_2$-PML, where $\epsilon_2\leq \epsilon_1$ is taken from $$\exp(-\epsilon_2)=\inf\limits_{\scriptsize \substack{\bm o\in\operatorname{supp}\left(\!P_O\!\right) \\ f'\in\operatorname{supp}\left(\!P_F\!\right)}}\mathbb{E}_{f\sim F}\left[\frac{P_{O|F=f}(\bm o)}{P_{O|F=f'}(\bm o)}\right].$$

   {\color{black}(2) Further, suppose  $\zeta_i^k\sim Lap(0,M^k)$. 
 If $\sum\limits_{k=0}^T\frac{\Delta''(k)}{M^k}\leq \epsilon_1,$
then algorithm $\mathcal{A}$ is $\epsilon_1$-DP for groups  of size $N$  and $\epsilon_2$-PML, where $\epsilon_2\leq \epsilon_1$ is taken from $\exp(-\epsilon_2)=\inf\limits_{\scriptsize \substack{\bm o\in\operatorname{supp}\left(\!P_O\!\right) \\ f'\in\operatorname{supp}\left(\!P_F\!\right)}}\mathbb{E}_{f\sim F}\left[\exp(-\sum\limits_{k=0}^T\frac{\Delta(k,f,f',\bm o)}{M^k})\right]$.}

\end{Theorem}
\vspace{3pt}

Theorem \ref{th::compare_DP_PML_upper_bound} demonstrates that the PML framework provides a tighter privacy guarantee than the DP framework in the entire view. 
Under any noise-adding mechanism, the PML framework enables algorithm designers to derive a more precise privacy leakage guarantee, $\epsilon_2$, compared to the DP guarantee $\epsilon_1$, where $\epsilon_2\leq\epsilon_1$. This distinction becomes evident when applying Laplace noise with computable privacy guarantees. {\color{black} PML strictly outperforms DP when, for all $\bm o \in \operatorname{supp}(P_O)$,
$\mathbb{E}_{f'\sim F}[P_{O|F=f'}(\bm o)]
>
\inf_{f'\in\operatorname{supp}(P_F)} P_{O|F=f'}(\bm o)$, which implies that the expectation under the prior is higher than the worst-case likelihood.}
Additionally, once the NE computation has DP guarantees, then it corresponds to a PML guarantee.  On the contrary, certain algorithms and game scenarios may be analyzable under the PML framework, but defy rigorous assessment with the DP framework. Specifically, there may be a privacy-preserving NE computation algorithm such that  $\inf\limits_{\bm o\in\operatorname{supp}\left(P_O\right)}\inf\limits_{f'\in\operatorname{supp}\left(P_F\right)}\mathbb{E}_{f\sim F}\left[\frac{P_{O|F=f}(\bm o)}{P_{O|F=f'}(\bm o)}\right]\geq \exp(-\epsilon)$ but $\sup\limits_{\bm o\in\operatorname{supp}\left(P_O\right)}\sup\limits_{f,f'\in\operatorname{supp}\left(P_{F}\right)}\frac{P_{O|F=f}(\bm o)}{P_{O|F=f'}(\bm o)}=\infty$ under the same condition. {\color{black}Notably, Theorem \ref{th::compare_DP_PML_upper_bound} develops the result under finite sets and one-step mapping \cite[Proposition 6]{saeidian2023pointwise} to general sets and iterative algorithms. 
Consequently, the PML refines DP from the entire view,  enabling flexibility for algorithm designers to better navigate tradeoffs among privacy guarantees, accuracy, regret, and other considerations \cite{ye2021differentially,wang2024differentially,lin2023statistical}.}

\vspace{-10pt}

\subsection{In Individual View}

{\color{black}We now examine the privacy leakage from each individual dataset, which leads to the comparison between the PML framework and the DP framework from an \textit{individual} view}. 
As outlined in \cite{saeidian2023pointwise,saeidian2024evaluating,saeidian2023rethinking}, the PML from an individual dataset $F_i$ is defined by 
$$\sup\limits_{\bm o\in\operatorname{supp}(P_O)}\max\limits_{i\in[N]}\ell(F_i\to \bm o).$$
It is important to note that datasets may exhibit either independence or correlation. We first consider the case of an independent dataset, where the variables $F_i$ and $F_{-i}$ are independent for any $i\in[N]$. Here $F_{-i}=\{F_1,\dots,F_{i-1},F_{i+1},\dots,F_N\}$ represents the set of variables corresponding to all players except player $i$. Similar to the definition of  $P_F$, let $P_{F_i}$ denote the distribution of $F_i$ and $P_{F_{-i}}$ denote the distribution of $F_{-i}$. Define
$\mathcal{Q}_{\mathcal{F}}=\{P_F|P_F=\Pi_{i=1}^{N}P_{F_i}\}$ as the set of all possible independent prior distributions. The relationship between PML in the individual perspective and DP in finite sets is shown in the following  \cite[Theorem 4.2]{saeidian2023rethinking}.

\begin{Proposition}[{\cite[Theorem 4.2]{saeidian2023rethinking}}]\label{pro::PML_DP_individual}
     Given $\epsilon$, algorithm $
     \mathcal{A}$ is $\epsilon$-DP if and only if
$\sup\limits_{P_F \in \mathcal{Q}_{\mathcal{F}}}\sup\limits_{\bm o\in\operatorname{supp}(P_O)}   \max\limits_{i \in[N]} \ell\left(F_i \rightarrow \bm o\right) \leq \epsilon$.
\end{Proposition}

{\color{black}The above proposition shows that the worst-case individual PML is equivalent to standard DP under independent priors, and for any given independent prior, the individual PML also provides a tighter guarantee than DP.}
The results differ significantly when datasets are correlated.  To make the following discussion more intuitive, we suppose that 
$\Omega_i=\mathbb{R}^n$.  Following \cite{saeidian2023pointwise,saeidian2023rethinking},  let $\epsilon_{\max }\left(F_i\right)\triangleq\log \frac{1}{\inf\limits_{f \in \operatorname{supp}(P_F)} P_{F_i}(f)}$ denote the largest amount of the information leakage about $F_i$ through any mechanism. This means we have $\ell (F_i\to \bm o)\leq \epsilon_{\max }\left(F_i\right)$ for all  $\bm o \in \mathcal{O}$ and $i\in[N]$. Additionally, to assess privacy guarantees for individual datasets, we provide a formulation for $\frac{P_{O|F_i=f_i,F_{-i}=f_{-i}}(\bm o)}{P_{O|F_i=f_i',F_{-i}=f_{-i}}(\bm o)}$ as stated in the following lemma. The proof is presented in Appendix \ref{ap::le::eq_different}.

\begin{Lemma}\label{le::eq_different}
Suppose $\zeta_i^k\sim Lap(0,M^k)$. For any $i\in[N]$, any possible $f_i,$ $f_i'$, $f_{-i}$ and $\bm o $,
we have $\frac{P_{O|F_i=f_i,F_{-i}=f_{-i}}(\bm o)}{P_{O|F_i=f_i',F_{-i}=f_{-i}}(\bm o)}
         =\frac{\int_{(\bm x'\!,\bm v')\in A(\bm o\!,\!\{f_i'\!,f_{\!-\!i}\!\}\!)}\! \Pi_{k\!=\!0}^TQ_i^k(\bm o,\bm v,\bm v', M^k)\Pi_{j\!=\!1}^NP_{O_j^{k}|F\!=\!f'\!,V_j^k\!=\!v_j^{\prime k}}(\!o_j^k\!)d \!(\!\bm x'\!,\bm v'\!)}{\int_{(\bm x',\bm v')\in A(\bm o,\{f_i',f_{-i}\})}\Pi_{k=0}^T\Pi_{j=1}^NP_{O_j^{k}|F=f',V_j^{k}=v_j^{\prime k}}(o_j^k)d(\bm x',\bm v')}\!,    $
         where $(\bm x,\bm v)\in A(\bm o,f)$.
\end{Lemma}

On this basis, we can find a lower bound for a special correlated dataset in the following theorem, whose proof is shown in Appendix \ref{ap::th::determinsistic game lower bound}.
\begin{Lemma}\label{th::determinsistic game lower bound}
Consider a case where each player has two possible cost functions, i.e., $f^0,$ and $f^1$, where the prior $P_F$ is generated by $P_{F_1}(f^0)=1-P_{F_1}(f^1)=\alpha$ and
\begin{equation}\label{eq::th3::P_F{-1}}
    \begin{aligned}
& P_{F_{-1} \mid F_1=f^1}\left(f_{-1}\right)= \begin{cases}\beta, & \text { if } f_{-1}=(f_1)^N, \\
\frac{1-\beta}{2^{N-1}-1}, & \text { otherwise } ,\end{cases}
\end{aligned}
\end{equation}
where  $0<\alpha<0.5$ and $0<\beta<1$. Suppose that $\zeta_i^k\sim Lap(0,M^k)$ and the algorithm $\mathcal{A}$ is $\epsilon$-DP. Then,

(1) $\sup\limits_{\bm o\in\operatorname{supp}\left(P_O\right)}\exp(\ell(F_1\to \bm o))\leq\frac{1}{(1\!-\!\alpha)\exp(-N\epsilon)+\!\alpha}.$



(2) Besides, there exists $\bm o \in\operatorname{supp}\left(P_O\right)$ such that $0<\epsilon_o\leq\epsilon$ and a lower bound of the PML from $F_1$ to $\bm o$ is 
$$     \begin{aligned}
& \exp(\ell(F_1\to \bm o))\\
    \geq&\frac{1}{(1\!-\!\alpha)\frac{\exp(\!-\!\epsilon_o)\left(\!\left(2^{N\!-\!1}\!-\!1\!\right)\exp(\!-\!\epsilon_o)^{N\!-\!1}\!+\!(1\!-\!\beta)\left(\!1\!+\!\exp(\!-\!\epsilon_o)\!\right)^{N\!-\!1}\!\right)}{\left(2^{N\!-\!1}\!-\!1\right)\!+\!(1\!-\!\beta)\left(\!1\!+\!\exp(\!-\!\epsilon_o)\!\right)^{N\!-\!1}}\!+\!\alpha}.
\end{aligned}$$

(3) Further, if  $\beta=\beta(N)$ is a function of $N$ such that $0<\beta(N)<1$, then
$\lim_{N\to \infty}\exp(\ell(F_1\to \bm o))=\frac{1}{\alpha}.$

\end{Lemma}
\vspace{3pt}

{\color{black}Lemma \ref{th::determinsistic game lower bound} establishes an upper bound and a lower bound for the PML guarantee in the individual view under a correlated dataset. }Noticing the prior $P_F$, we consider a case where the cost functions of other players are likely to align with the $1$st player's cost function with a probability $\beta$.{\color{black} Compared to one-step privacy mechanisms and the given constant $\beta$ in \cite[Theorem 3]{saeidian2024evaluating}, this work develops the analysis to a widely used iterative algorithm in aggregative games and considers any probability $\beta(N)$ that depends on the number of players.}

Further, based on the above result, we can compare a lower bound of PML guarantees in the individual view and an upper bound of DP guarantees as shown in the following theorem,  the proof of which could be found in Appendix \ref{ap::th::compare_DP_PML_lower_boundd}.

\begin{Theorem}\label{th::compare_DP_PML_lower_boundd}
Suppose $\zeta_i^k\sim Lap(0,M^k)$.
    For any $\epsilon_1$ and $\epsilon_2\in (0,\max\limits_{i\in[N]}\epsilon_{\max }(F_i))$, we can construct correlated datasets of players' cost functions and algorithm $\mathcal{A}$ such that 
    
    (1) algorithm $\mathcal{A}$ is $\epsilon_1$-DP, and 
    
    (2) $\sup\limits_{\bm o  \in \operatorname{supp}\left(P_O\right)}\max\limits_{i\in[N]}\ell (F_i\to \bm o)\geq\epsilon_2.$
\end{Theorem}

{\color{black}Theorem \ref{th::compare_DP_PML_lower_boundd} demonstrates that it is possible to construct specific datasets of players' cost functions and algorithms with arbitrary DP guarantees and arbitrary lower bounds for PML guarantees from the individual perspective.} Notably, if we set $\epsilon_1<\epsilon_2$ in Theorem \ref{th::compare_DP_PML_lower_boundd},  the lower bound of the PML guarantee in the individual view exceeds the upper bound provided by the DP guarantee.  This difference becomes particularly obvious when $\epsilon_1$ approaches $0$ and $\epsilon_2$ approaches $\max\limits_{i\in[N]}\epsilon_{\max }(F_i)$. {\color{black}As a result, when the real privacy leakage of correlated datasets is significant and high, the PML framework offers a proper assessment of privacy leakage, while the DP framework fails to capture the privacy leakage. We are the first to compare their performance for the privacy-preserving NE computing algorithms in aggregative games, which underscores the importance of replacing the DP framework with the PML framework for assessing privacy leakage, particularly in the presence of correlated datasets.}


{\color{black} 
\begin{remark}
    In this section, we clarify the relation between the  PML guarantee and the DP guarantee. Specifically, as shown in Theorem 2,  the PML guarantee is tighter than the DP guarantee in the entire view. Besides, under the individual view, the PML guarantee, considering the independent datasets, is tighter than the DP guarantee (Proposition 3), while considering the correlated dataset,  the lower bound of PML may exceed the upper bound of DP (Theorem 3). 
\end{remark}
\begin{remark}
     Note that we compare PML with DP in the entire view, as well as compare the individual PML with DP, since they are in the same privacy level. Specifically, $\epsilon$-DP for groups of size $N$ may correspond to $\frac{\epsilon}{N} $-DP at the individual level according to Proposition 2. Besides, the PML satisfies the same composition property $\ell(F \rightarrow \boldsymbol{o}) \leq \sum_{i=1}^N \ell\left(F_i \rightarrow \boldsymbol{o}\right)$, where $F= \left\{F_1, \ldots, F_N\right\}$. Consequently, PML $\ell(F \rightarrow \boldsymbol{o})$ is the natural counterpart to DP for groups of size $N$, as both characterize the privacy of the entire dataset, while individual PML $\ell\left(F_i \rightarrow \boldsymbol{o}\right)$ is the appropriate counterpart to the  DP defined over adjacent datasets.
\end{remark}

\begin{remark}\label{rem:comparison_R1}
Different from the non-cooperative games in \cite{chen2025novel}, our framework is developed for typical aggregative games. In aggregative games, each player's cost depends on the aggregate term $\delta(x)=\frac{1}{N}\sum_{i=1}^N x_i$, which is not directly observable and must be estimated through shared noisy observations $o_i^k=v_i^k+\zeta_i^k$. This leads to a different sensitivity characterization compared with non-cooperative games without aggregation. Besides, the analysis in \cite{chen2025novel} focuses on an averaged measure, i.e., $\frac{1}{N}\sum_{i=1}^N \ell(F_i\to \bm o)$, whereas our work provides a finer characterization, including the PML $\ell(F\to \bm o)$ and the individual PML $\ell(F_i\to \bm o)$. Moreover, we  first establish that in games with correlated datasets, the individual PML can exceed the corresponding DP guarantees. 

\end{remark}

}

\section{Numerical Experiments}


In this section, we provide privacy-preserving simulations in a disease problem \cite{kifer2011no,bauch2004vaccination} with the adversary's guess \cite{zhu2019deep} to show the effectiveness of our results.

\vspace{-10pt}

\subsection{Game Model and Privacy-Preserving Algorithm}

We focus on an aggregative game with $N$ players exposed to contagious diseases, and players decide whether to vaccinate \cite{kifer2011no,bauch2004vaccination}. Consider the model with multi-dimensional strategy spaces wherein each player is endowed with $n$ distinct vaccination alternatives. The strategy of the $i$th player, $x_i$, represents a probability distribution over these alternatives (i.e., each entry in $x_i$ corresponds to the likelihood of choosing a particular vaccine). Let $A_i$ and $B_i$ denote the morbidity risks for vaccination and non‐vaccination, respectively, for the $i$th player without considering other players.  Besides, since unvaccinated players are influenced by the vaccination choices of others \cite{bauch2004vaccination}, we introduce $\delta(x)$ to represent the vaccination coverage status among players.  Consequently, the cost function for the $i$th player is defined as: $f_i(x_i,x_{-i})=A_ix_i+\delta(x)^TB_i(\mathbf{1}_n-x_i)$, where $\mathbf{1}_n$ is the $n$-dimensional column vector with all entries equal to $1$. {\color{black}This formulation satisfies the structure of an aggregative game, as it can be rewritten as $\bar{f}_i(x_i, \delta(x))$, where the cost depends only on player $i$'s own strategy $x_i$ and the aggregate term $\delta(x)$. }Further, we incorporate a prior probability model capturing the infection risk among players \cite{kifer2011no,saeidian2023rethinking}. Following the prior distribution in Lemma \ref{th::determinsistic game lower bound}, let $\alpha$ represent the baseline infection probability for the $1$st player, and $\beta$ represent the correlation coefficient showing how the $1$st player's infection status affects others. The explicit formulation of this prior is given in (\ref{eq::th3::P_F{-1}}). Notably, the infection status of each player influences the morbidity risks, $A_i$ and $B_i$. For notational convenience, we define $A=[A_1,\dots,A_N]$ and $B=[B_1,\dots,B_N]$ as the parameter matrices of the game. Take $\theta=[A(1)^T,\dots,A(N)^T,B(1)^T,\dots,B(Nn)^T]^T$ as the parameter vector generated by matrix $A$ and $B$, where $A(i)$ is the $i$th column of matrix $A$ and  $B(j)$ is the $j$th column of matrix $B$, for $i\in[N]$ and $j\in[Nn]$. By \cite{ye2021differentially}, the strategy updating for the $i$th player at iteration $k$ is given by $x_i^{k+1}=x_i^k-\lambda^k F_i(x_i^k,v_i^k)$, while the estimated aggregated term is updated as $v_i^{k+1}=\sum_{j=1}^N\frac{1}{N}o_j^k +x_i^{k+1}-x_i^k.$ Similar to settings in \cite{wang2024ensuring}, we take $M^k=p_1+p_2k^{p_3}$ as the variance in Laplace distribution and $\lambda^k=\frac{q_1}{q_2+q_3k}$ as the stepsize, where $p_i,q_i\in[0,2]$.

  \vspace{-5pt}
\subsection{Adversary Model and Gain Function}

Consider that the adversary attempts to guess private information by leveraging the algorithm inspired by \cite{zhu2019deep}. Specifically, in our model, the adversary aims to infer the parameter vector $\theta$ and constructs a guess $\hat{\theta}$. Take $\hat{A}$ and $\hat{B}$ as the predicted values of $A$ and $B$, respectively.  
{\color{black}The adversary's objective is to maximize the alignment between its guess $\hat{\theta}$ and the true value $\theta$, by simulating the algorithmic process based on observed information through gradient matching \cite{zhu2019deep}. }Specifically,  the adversary initializes dummy matrices $\hat{A}_i^1$ and $\hat{B}_i^1$ by $\mathcal{N}(0,1)$ and the dummy strategy by $\hat{x}_i^0=o_i^0$. At iteration $k$, the adversary computes the pseudo-gradient for the $i$th player as $\hat{F}_i^k(x_i,\delta(x))= \nabla_{x_i} \left\{\hat{A}_i^kx_i+\delta(x)^T\hat{B}_i^k(\mathbf{1}_n-x_i)\right\}. $ The adversary simulates the strategy update by $\hat{x}_i^{k+1}=\hat{x}_i^k-\lambda^k \hat{F}_i(\hat{x}_i^k,\hat{v}_i^k)$ and  the aggregate update by $\hat{v}_i^{k+1}=\sum_{j=1}^N\frac{1}{N}o_j^k +\hat{x}_i^{k+1}-\hat{x}_{i}^k.$ Then the adversary minimizes the distance between the dummy aggregate term $\hat{v}_i^{k+1}$ and the observation $o_i^{k+1}$, i.e., $$\mathbb{D}_i^k=\|\hat{v}_i^{k+1}-o_i^{k+1}\|^2=\|\sum_{j=1}^N\frac{1}{N}o_j^k +\lambda^k\hat{F}_i^k(\hat{x}_i^k,o_i^k)-o_i^{k+1}\|^2,$$ and update  $\hat{A}_i^{k+1}, \hat{B}_i^{k+1}$ via gradient descent with stepsize $\nu$. Inspired by \cite{zhu2019deep}, with dummy strategy $\hat{x}_i^k$, the algorithm of the adversary for each $i$ is summarized in the following
\begin{equation}\label{alg::guess}
\begin{cases}
\mathbb{D}_i^k=\|\sum\limits_{j=1}^N\frac{1}{N}o_j^k +\lambda^k\hat{F}_i^k(\hat{x}_i^k,o_i^k)-o_i^{k+1}\|^2,\\
\hat{A}_i^{k+1}=\hat{A}_i^{k}-\nu  \nabla_{\hat{A}_i^{k}}\mathbb{D}_i^k,\\
\hat{B}_i^{k+1}=\hat{B}_i^{k}-\nu  \nabla_{\hat{B}_i^{k}}\mathbb{D}_i^k.
\end{cases}
\end{equation}
Finally, after $T$ iterations, the algorithm outputs $\hat{A}_i^{T+1}$ and $\hat{B}_i^{T+1}$, and the adversary obtains the final guess $\hat{\theta}$.

{\color{black}Further,  using the true parameter vector $\theta$ and the adversary's guess vector $\hat{\theta}$, we construct a gain function based on the cosine similarity. Specifically, we define 
$g(\theta,\hat{\theta})=\frac{\theta^T\hat{\theta}}{\|\theta\|\|\hat{\theta}\|}+1,$ 
 where the first term is the cosine similarity between the true parameter vector $\theta$ and the adversary’s estimated vector $\hat{\theta}$. Notice that the addition of $1$ ensures that  $0\leq g(\theta,\hat{\theta})\leq 2$ and makes $g\in\Gamma$. This formulation provides clear quantitative interpretation within the $g$-leakage method \cite{m2012measuring}:  When $g(\theta,\hat{\theta})$ approaches $2$ (i.e., cosine similarity approaches 1), it indicates strong alignment between the adversary's guess and the true parameters, suggesting high privacy leakage; conversely, when the value of $g(\theta,\hat{\theta})$ approaches $1$, it reflects poor inference accuracy and low privacy leakage. Thus, the gain function is able to assess both the adversary's inference capability and the algorithm's privacy-preserving performance.}


\begin{figure}[t]
\vspace{-10pt}
    \centering
    \includegraphics[width=0.35\textwidth]{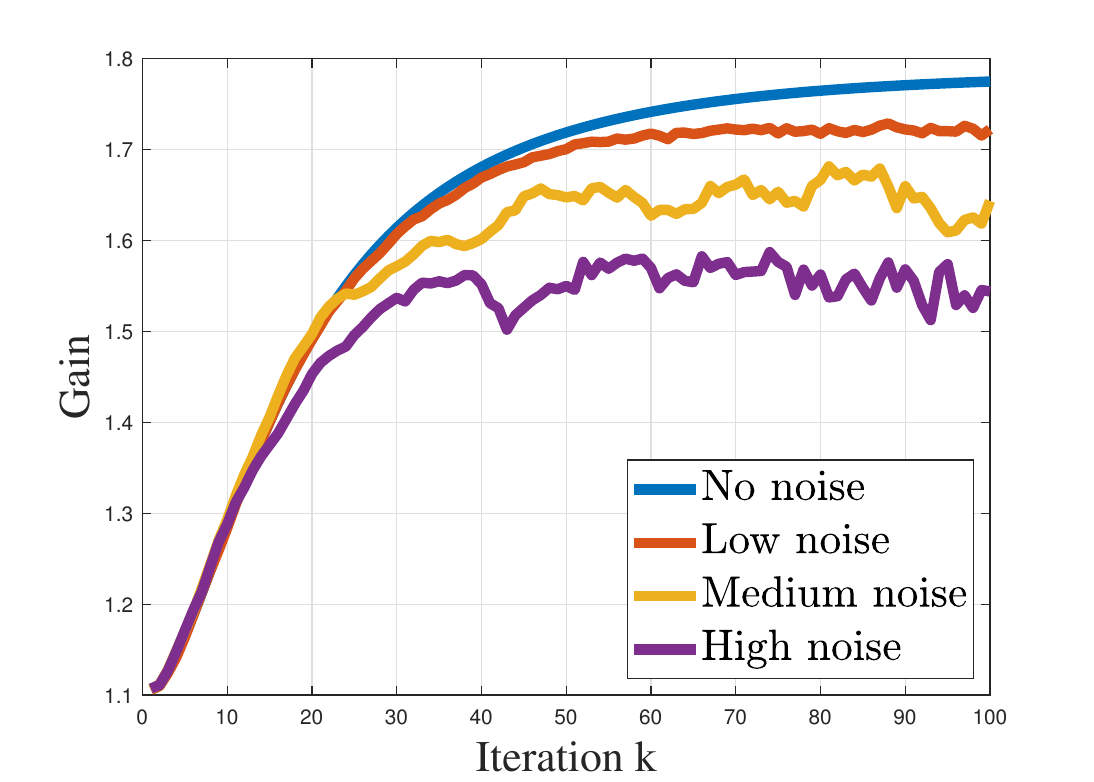}
    \caption{Value of  $g(\theta,\hat{\theta})$ under different noise levels. The blue line, red line, yellow line, and purple line represent the adversary's gain values when algorithm $\mathcal{A}$ incorporates no noise, low noise ($M^k=1+0.5k$), medium noise ($M^k=1+k$), and high noise ($M^k=1+1.5k$), respectively. }
    \label{fig::gain}
    \vspace{-15pt}
\end{figure}

Fig. \ref{fig::gain} verifies the effectiveness of the adversary's guess algorithm (\ref{alg::guess}). 
Specifically, the gain value increases with iterations and asymptotically approaches a neighborhood of $2$, especially in the case with no noise. This empirical evidence suggests that the adversary's guess algorithm successfully converges toward a neighborhood of the true parameter vector $\theta$. Furthermore, the higher noise brings a lower gain value. This observation aligns with our privacy preservation objective, as increased noise variance in algorithm $\mathcal{A}$ reduces information leakage, thereby enhancing privacy preservation.
\subsection{Comparison between PML and DP}

In the following, we compare the privacy guarantees of PML and DP, and assess the real leakage performance of the privacy-preserving algorithm proposed in \cite{ye2021differentially} when confronted with the adversary described in \cite{zhu2019deep}. Notice that, without the observation, the adversary cannot infer a meaningful $\hat{\theta}'$ correlated with $\theta$, resulting in a baseline gain value $g(\theta,\hat{\theta}')=1$. Recall that $g(\theta,\hat{\theta})$ is  value of the gain function by observes $\bm o$, where $\hat{\theta}$ is the output of guess algorithm  (\ref{alg::guess}). This allows us to express the gain ratio as: $\log(g(\theta,\hat{\theta}))=\log\left(\frac{g(\theta,\hat{\theta})}{1}\right)=\log\left(\frac{g(\theta,\hat{\theta})}{g(\theta,\hat{\theta}')}\right)$. 
Recalling the formulation of $\ell(F \rightarrow \bm o)$ in (\ref{eq::PML}), it is reasonable for us to use the  $\log(g(\theta,\hat{\theta}))$ as the real information leakage, when facing the adversary in \cite{zhu2019deep}. In our subsequent simulation, the value of $g(\theta,\hat{\theta})$ is taken by the average of $100$ cases generated by the prior distribution. 

Fig. \ref{fig::entire dataset} compares three key indexes: PML guarantees in Theorem \ref{th::determinsistic game upper bound}, DP guarantees for groups of size $N$, and the real information leakage when facing the adversary in \cite{zhu2019deep} with the gain function. Notice that the upper bound of the PML guarantees is lower than that of the DP guarantees for groups of size $N$.  It is consistent with Theorem \ref{th::compare_DP_PML_upper_bound} that the PML framework provides tighter privacy guarantees compared to DP in the entire view.  {\color{black} Besides, since the values of $\log(g(\theta,\hat{\theta}))$ measure the real information leakage in this case,  the growth of the leakage is below the privacy guarantee in the PML framework. It verifies that PML effectively captures worst-case information leakage scenarios as shown in Definition \ref{def::PML} and Theorem \ref{th::determinsistic game upper bound}. }Therefore, we provide robust and operational privacy guarantees by the PML framework, which refines the over-conservative assessment in the DP framework. 

{\color{black}
Since the tradeoff between accuracy and privacy is very important, we aim to compare the performance of PML and DP in the following.  According to (\ref{eq::th::PML}), the PML guarantee is characterized by $\epsilon_{PML}(\bm \lambda,\bm M)$, where $\bm \lambda=\{\lambda^1,\dots,\lambda^T\}$ and $\bm M =\{M^1,\dots,M^T\}$.  Take mean squared error at iteration $T$ as $\mathbb{E}(\|x^T-x^*\|^2)$, and the designer considers the following optimization problem:
$\min_{\bm\lambda,\bm M} \ \mathbb{E}(\|x^T-x^*\|^2)$ such that 
$\epsilon_{PML}(\bm \lambda,\bm M) \leq R$, since the privacy leakage should not exceed this budget. Under the DP framework, the constraint becomes
$\epsilon_{DP}(\bm \lambda,\bm M) \le R.$ By considering the same parameter settings as in \cite{ye2021differentially},
we compare the errors under  PML and DP constraints, respectively, in Fig. \ref{fi::2-1} (a).  Moreover, by adopting the aggregative estimation structure in \cite{wang2024differentially}, we further compare the error under the same privacy budget, as illustrated in Fig.~\ref{fi::2-1}(b). Noting that a lower error means a  higher accuracy, the results show that algorithms under the PML guarantee achieve higher accuracy and enable better flexibility for better tradeoffs.



}

\begin{figure}[tbp]
\vspace{-13pt}
    \centering
    \includegraphics[width=0.37\textwidth]{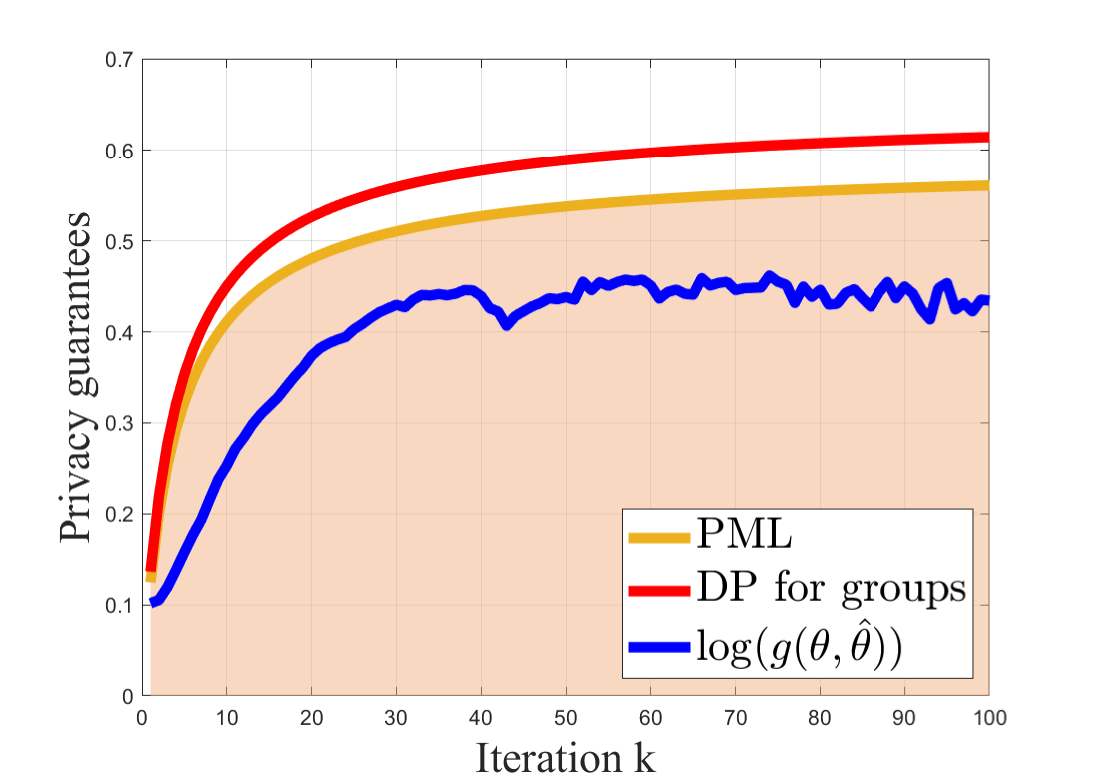}
    \caption{Comparison between PML and DP in the entire view. The orange line and the red line show the upper bound of the PML guarantees and the DP guarantees for groups, respectively.  The blue line denotes the real information leakage when facing the adversary in \cite{zhu2019deep} with the gain function.}
    \label{fig::entire dataset}
    \vspace{-15pt}
\end{figure}
\begin{figure}[tbp]
\vspace{-10pt}
 \centering
\subfigure[Tradeoff comparison with Ye et al. \cite{ye2021differentially}]{
 \includegraphics[width=2.8in]{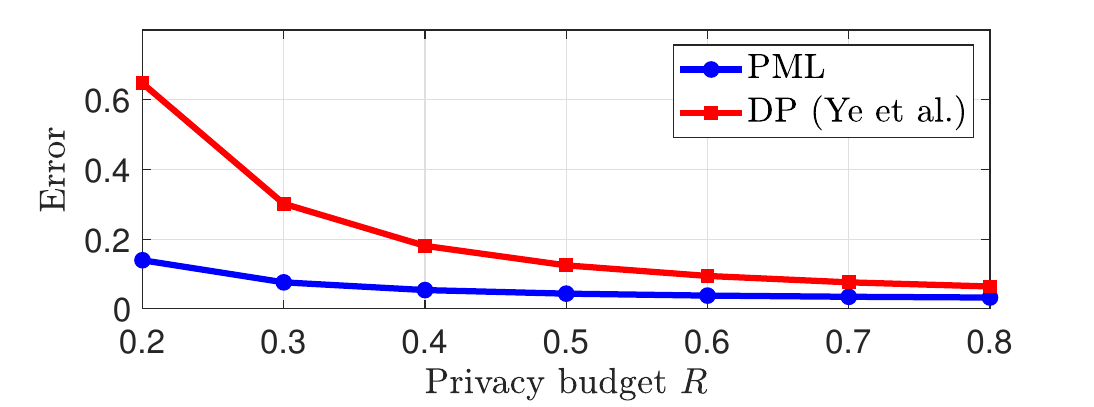}
 \vspace{-20pt}
 }\\
 \subfigure[Tradeoff comparison with Wang et al. \cite{wang2024differentially}]{
 \includegraphics[width=2.8in]{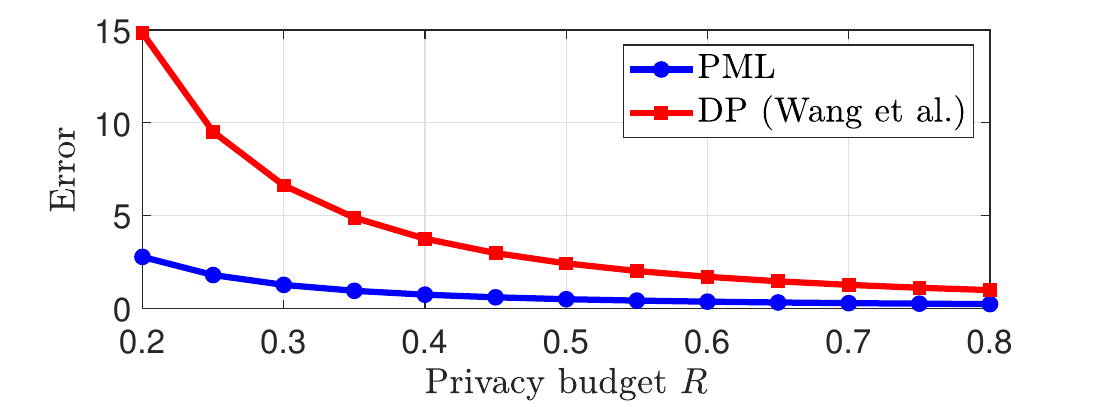}
 \vspace{-20pt}
 }
 \vspace{-5pt}
\caption{\color{black}Tradeoff comparison between error and privacy budget $R$ under PML and DP frameworks. The blue lines represent the tradeoff under the PML guarantee, while the red lines represent the DP-based results from \cite{ye2021differentially} and \cite{wang2024differentially}. 
}
\label{fi::2-1}
\vspace{-5pt}
\end{figure}

\begin{figure}[t]
    \centering
    \includegraphics[width=0.37\textwidth]{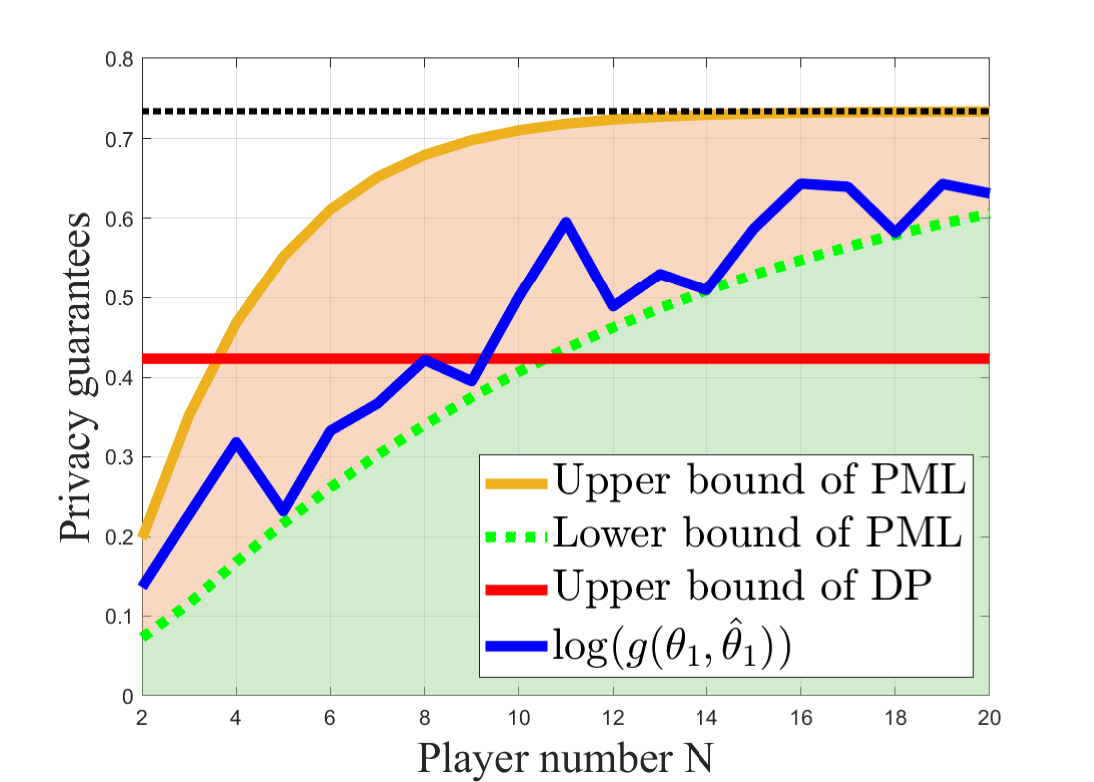}
    \caption{Comparison between PML and DP in the individual view. The red line and the green dot line represent an upper bound and a lower bound of PML guarantees in the individual view, according to Theorem \ref{th::compare_DP_PML_lower_boundd}. The red line describes the upper bound of the DP guarantees that remains constant, while correlation increases with the number of players. The black dotted line shows the upper bound of the privacy leakage,  $\epsilon_{\max }\left(F_1\right)$, which is approached by the upper and lower PML bounds. The blue line denotes the real information leakage of the $1$st player when facing the adversary in \cite{zhu2019deep} with the gain function. This experiment shows that DP is not able to capture the effect of correlations on the privacy risk.}
    \label{fig::related lower bound}
    \vspace{-17pt}
\end{figure}

Furthermore, analogous to $\log(g(\theta,\hat{\theta}))$, we take $\log(g(\theta_1,\hat{\theta}_1))$ as the real information leakage in the individual view.  As illustrated in Fig. \ref{fig::related lower bound}, we take the lower bound of individual PML guarantees in Lemma \ref{th::determinsistic game lower bound}, i.e., $\ell(F_1\to \bm o)$. Besides, the blue line surpasses the red line when $N\geq10$, aligning with Theorem \ref{th::compare_DP_PML_lower_boundd}. This confirms that the lower bound of the PML guarantees in the individual view may exceed the upper bound provided by DP guarantees. Noticing that we draw a lower bound of PML in an individual, there may exist cases where the real information leakage exceeds the green dot line, as shown in Fig. \ref{fig::related lower bound}. 
Notably, the blue line—representing the real 
information leakage—lies above the red line when
$N>10$, demonstrating that the real leakage can indeed surpass the DP upper bound. This discrepancy highlights a critical limitation of the DP framework in accurately quantifying information leakage in such cases.  These findings underscore the necessity of adopting the PML framework over DP for privacy leakage assessment, as PML provides a proper assessment of privacy risks and DP fails to assess the leakage in this situation.

\section{Conclusion and Discussion}

In this paper, we have proposed a PML framework for assessing the privacy guarantee of NE-computing algorithms in aggregative games.  Given prior knowledge, we have derived a precise and computable upper bound of privacy leakage with PML guarantees for NE-computing algorithms. Additionally, we have conducted a comparative analysis between PML and DP. In the entire view, we have demonstrated that the PML refined DP by offering a tighter privacy guarantee, enabling flexibility in designing NE computation. Also, in the individual view, we have revealed that the lower bound of PML can exceed the upper bound of DP by constructing specific correlated datasets. As a result, when the real privacy leakage of correlated datasets was significant, we have demonstrated that PML was a proper privacy guarantee, while DP failed to capture the privacy leakage. Finally, we have designed experiments involving adversaries attempting to infer private information to confirm the effectiveness of our method.

{\color{black}
Several privacy notions have been compared with PML in finite datasets, including mutual information, local DP, and so on \cite{saeidian2023pointwise,saeidian2026information}, where privacy notions were provided for different situations. For example, mutual information privacy \cite{wang2016relation} bounds average information leakage, while PML characterizes the leakage from a pointwise view of the observation, and is operational with outside adversary models \cite{saeidian2023pointwise, issa2019operational}. Local DP \cite{chen2024local} and Rényi DP \cite{mironov2017renyi} are evaluated for privacy mechanisms without knowledge of the prior distribution,  while our PML results bring a proper privacy guarantee to avoid the ineffective assessments for games with correlations by traditional DP. Thus, our PML framework is proper and operational for games with prior distributions or correlated datasets.

Moreover, since PML is defined over the prior distribution $P_F$, we have provided how to compute the information leakage based on PML under the prior distribution. If the distribution is not available, one may use the estimated distribution $P_F$ to design the algorithm. For instance, the recent empirical PML work \cite{grosse2025privacy} shows that one can derive distribution-independent PML guarantees via estimating the distribution from data samples. The extension for empirical PML by estimating the distribution in a distributed manner constitutes an interesting direction for future research. Moreover, while Theorem~\ref{th::compare_DP_PML_lower_boundd} considers a binary correlated dataset construction, extending the analysis to more general correlated datasets and general noise remains open. Extensions of the proposed framework to stochastic games, online games, and distributed optimization also constitute promising future directions.


}
}

\section*{Appendix}

\subsection{Proof of Lemma \ref{le::PMLconvert}}\label{ap::le::PMLconvert}

{\color{black}The proof consists of two steps. We first derive an upper bound on the  PML guarantee, and then we express this bound in expectation, which directly yields the sufficient condition.

\textbf{Step 1:} We first establish the bound of $\sup\limits_{\bm o\in\operatorname{supp}\left(P_O\right)}\!\!\exp \ell(F\to \bm o)$ by
\(
\frac{P_O(\bm o)}{P_{O|F=f'}(\bm o)}
\) considering all possible distribution. }For any $\bm o\in\operatorname{supp}\left(P_O\right)$, we have $P_O(\bm o)>0$ and there exists $f'\in\operatorname{supp}\left(P_F\right)$ such that $P_{O|F=f'}(\bm o)>0$. Recall the convention that $\frac{x}{0} =\infty$ if $x > 0$. For all 
$f'\in\operatorname{supp}\left(P_F\right)$ such that  $P_{O|F=f'}(\bm o)=0$, we have $\frac{P_{O}(\bm o)}{P_{O|F=f'}(\bm o)}=\infty$. Then $\inf\limits_{f'\in\operatorname{supp}\left(P_F\right),P_{O|F=f'}(\bm o)\neq0}\frac{P_{O}(\bm o)}{P_{O|F=f'}(\bm o)}<\inf\limits_{f'\in\operatorname{supp}\left(P_F\right),P_{O|F=f'}(\bm o)=0}\frac{P_{O}(\bm o)}{P_{O|F=f'}(\bm o)}=\infty$. 
Thus, for any $\bm o\in\operatorname{supp}\left(P_O\right)$, recalling that there exists $f'\in\operatorname{supp}\left(P_F\right)$ such that $P_{O|F=f'}(\bm o)\in(0,1]$,  we obtain $$\begin{aligned}
\!\frac{P_{O}(\bm o)}{\sup\limits_{f'\in\operatorname{supp}\left(P_F\right)}\!\!P_{O|F=f'}(\bm o)}
\!=&\frac{P_{O}(\bm o)}{\sup\limits_{f'\in\operatorname{supp}\left(P_F\right),P_{O|F=f'}(\bm o)\neq0}P_{O|F=f'}(\bm o)}\\
=&\inf\limits_{f'\in\operatorname{supp}\left(P_F\right),P_{O|F=f'}(\bm o)\neq0}\frac{P_{O}(\bm o)}{P_{O|F=f'}(\bm o)}\\
=&    \inf\limits_{f'\in\operatorname{supp}\left(P_F\right)}\frac{P_{O}(\bm o)}{P_{O|F=f'}(\bm o)}.
\end{aligned}$$
Besides, since $P_{O}(\bm o)>0$, we have
 $$\begin{aligned}
     \!\!  \sup\limits_{\bm o\in\operatorname{supp}\left(P_O\right)}\!\!\exp \ell(F\to \bm o) \!
       =&\sup\limits_{\bm o\in\operatorname{supp}\left(P_O\right)}\sup\limits_{f'\in\operatorname{supp}\left(P_F\right)}\frac{P_{O|F=f'}(\bm o)}{P_{O}(\bm o)}\\
       =&\sup\limits_{\bm o\in\operatorname{supp}\left(P_O\right)}\frac{1}{\frac{P_{O}(\bm o)}{\sup\limits_{f'\in\operatorname{supp}\left(P_F\right)}P_{O|F=f'}(\bm o)}}\\
       =&\sup\limits_{\bm o\in\operatorname{supp}\left(P_O\right)}\frac{1}{\inf\limits_{f'\in\operatorname{supp}\left(P_F\right)}\frac{P_{O}(\bm o)}{P_{O|F=f'}(\bm o)}}\\
       \leq&\frac{1}{\inf\limits_{\bm o\in\operatorname{supp}\left(P_O\right)}\inf\limits_{f'\in\operatorname{supp}\left(P_F\right)}\frac{P_{O}(\bm o)}{P_{O|F=f'}(\bm o)}}.
       \end{aligned}
       $$

{\color{black}\textbf{Step 2:} We next rewrite the ratio
$\frac{P_O(\bm o)}{P_{O|F=f'}(\bm o)}$
using the law of total probability. If $P_{O|F=f'}(\bm o)>0$, then
\begin{equation}\label{eq::pf::lem1::ven}
    \begin{aligned}
    \frac{P_{O}(\bm o)}{P_{O|F=f'}(\bm o)}
    =&\frac{\int_{f\in\operatorname{supp}\left(P_F\right)}P_{O|F=f}(\bm o)P_F(f)d f}{P_{O|F=f'}(\bm o)}\\
    =&\int_{f\in\operatorname{supp}\left(P_F\right)}\frac{P_{O|F=f}(\bm o)}{P_{O|F=f'}(\bm o)}P_F(f)d f.
\end{aligned}
\end{equation}
Besides, if $P_{O|F=f'}(\bm o)=0$, equation (\ref{eq::pf::lem1::ven}) also holds by the convention $\frac{x}{0}=\infty$, where $x>0$. }Thus,
    $$\begin{aligned}
       &\sup\limits_{\bm o\in\operatorname{supp}\left(P_O\right)}\exp \ell(F\to \bm o) \\
       \leq &\frac{1}{\inf\limits_{\bm o\in\operatorname{supp}\left(P_O\right)}\inf\limits_{f'\in\operatorname{supp}\left(P_F\right)}\int_{f\in\operatorname{supp}\left(P_F\right)}\frac{P_{O|F=f}(\bm o)}{P_{O|F=f'}(\bm o)}P_F(f)d f}\\
       =&\frac{1}{\inf\limits_{\bm o\in\operatorname{supp}\left(P_O\right)}\inf\limits_{f'\in\operatorname{supp}\left(P_F\right)}\mathbb{E}_{f\sim F}\left[\frac{P_{O|F=f}(\bm o)}{P_{O|F=f'}(\bm o)}\right]}
       \end{aligned}
       $$
Therefore, if $\inf\limits_{\bm o\in\operatorname{supp}\left(P_O\right)}\inf\limits_{f'\in\operatorname{supp}\left(P_F\right)}\mathbb{E}_{f\sim F}\left[\frac{P_{O|F=f}(\bm o)}{P_{O|F=f'}(\bm o)}\right]\geq \exp(-\epsilon)$, then $\sup\limits_{\bm o\in\operatorname{supp}\left(P_O\right)}\ell(F\to \bm o) \leq \epsilon$. $\hfill\square$

\subsection{Proof of Lemma \ref{le::1}}\label{ap::le::1}

{\color{black}The proof proceeds in two steps. Step 1 establishes an integral representation of
$P_{O|F=f}(\bm o)$. Step 2 expresses the likelihood ratio $\frac{P_{O|F=f}(\bm o)}{P_{O|F=f'}(\bm o)}$ by exploiting the Laplace mechanism to obtain the desired expression.

\textbf{Step 1:} We first express the conditional observation distribution $P_{O|F=f}(\bm o)$
by integrating over all feasible state and aggregate trajectories.} At iteration $k$, recall that $O^k$ and $V^k$ are the random variables representing the observation $o^k$ and the estimate term $v^k$ at iteration $k$, respectively. We further take $X^k$ as the random variable representing the strategy $x^k$. Take $O^{[k]}$, $X^{[k]}$, and $V^{[k]}$ as the random variables representing the observation $\{\bm o^l\}_{l=0}^k$, the strategy $\{x^l\}_{l=0}^k$, and estimate term $\{v^l\}_{l=0}^k$  from iteration $l=0$ to iteration $k$, respectively. To avoid ambiguity,  if $l=0$, take $\{o^{l-1}, x^l,v^l\}=\{ x^l,v^l\}$ and $O^{[l-1]}=\varnothing$. Similar to \cite{huang2015differentially}, for any possible $\bm o ,$ $f$, and $T$, 
$$
\begin{aligned}
&P_{O|F=f}(\bm o)\\
    =&\!\!\!\!\int\!\!P_{O^{[T]}\!,X^{[T]}\!,V^{[T]}|F\!=\!f}(\{o^k\!, x^{k}\!, v^{k}\}_{k\!=\!0}^T\!)d(\!\bm x\!,\!\bm v\!)\\
    =&\!\!\!\!\int\!\!\!\Pi_{k=0}^{T}\{ P_{O^{k}|F=f,\{O^{[k-1]},X^{[k]},V^{[k]}\}=\{o^{l-1},x^l,v^l\}_{l=0}^{k}}(o^k)\\
    & \quad \cdot P_{X^{k}\!,V^{k}|F\!=\!f,\{O^{[k\!-\!1]}\!,X^{[k\!-\!1]}\!,V^{[k\!-\!1]}\}\!=\!\{o^{l}\!,x^l\!,v^l\}_{l\!=\!0}^{k\!-\!1}}\!( x^{k}\!, v^{k})\}d(\!\bm x\!,\!\bm v\!).
\end{aligned}
$$
Notice that for $i\in [N]$, $o_i^k=v_i^k+\zeta_i^k$, and for $j\neq i$, $\zeta_i^k$ and $\zeta_{j}^k$ are independent. Then,
$\Pi_{i=1}^NP_{O_i^{k}|F=f,V_i^k=v_i^k}(o_i^k)=P_{O^{k}|F=f,\{O^{[k-1]},X^{[k]},V^{[k]}\}=\{o^{l-1},x^l,v^l\}_{l=0}^{k}}(o^k).
$
For any $j\neq i$, $v_i^{k}$ and $v_{j}^{k}$ are independent given any $x^k, x^{k-1}, v^{k-1}$, and $o^{k-1}$, while $x_i^{k}$ and $x_{j}^{k}$ are independent for given any $ x^{k-1}, v^{k-1},o^{k-1},$ and $ f$. Then
$$
\begin{aligned}
    & P_{X^{k}\!,V^{k}|F\!=\!f,\{O^{[k\!-\!1]}\!,X^{[k\!-\!1]}\!,V^{[k\!-\!1]}\}\!=\!\{o^{l}\!,x^l\!,v^l\}_{l\!=\!0}^{k\!-\!1}}\!( x^{k}\!, v^{k})\\
    =& \Pi_{i=1}^NP_{X_i^{k}|F\!=\!f,\{O^{[k\!-\!1]}\!,X^{[k\!-\!1]}\!,V^{[k\!-\!1]}\}\!=\!\{o^{l}\!,x^l\!,v^l\}_{l\!=\!0}^{k\!-\!1}}\!(x_i^k\!)\\
    &\Pi_{i=1}^NP_{V_i^{k}|F\!=\!f,\{O^{[k\!-\!1]}\!,X^{[k\!-\!1]}\!,V^{[k\!-\!1]}\}\!=\!\{o^{l}\!,x^l\!,v^l\}_{l\!=\!0}^{k\!-\!1},X^k\!=\!x^k}\!(v_i^k).
\end{aligned}$$
Additionally, according to algorithm $\mathcal{A}$, $P_{V_i^{k}|F\!=\!f,\{O^{[k\!-\!1]}\!,X^{[k\!-\!1]}\!,V^{[k\!-\!1]}\}\!=\!\{o^{l}\!,x^l\!,v^l\}_{l\!=\!0}^{k\!-\!1},X^k\!=\!x^k}\!(v_i^k)=1$, if and only if $v_i^{k}=h_v(v_i^{k-1},x_i^{k},x_i^{k-1},o^{k-1},\gamma^{k-1})$, and $P_{X_i^{k}|F\!=\!f,\{O^{[k\!-\!1]}\!,X^{[k\!-\!1]}\!,V^{[k\!-\!1]}\}\!=\!\{o^{l}\!,x^l\!,v^l\}_{l\!=\!0}^{k\!-\!1}}\!(x_i^k\!)=1$, if and only if $x_i^{k}=h_x(x_i^{k-1},v_i^{k-1},\lambda^{k-1},f_i)$.
Recall the definition of $A(o^{k-1},f| x^{k-1}, v^{k-1})$ in (\ref{eq::A(ofxvk)}) and $A(\bm o, f)$ in (\ref{eq::Aof}), we have
$$
\begin{aligned}
    P_{O|F\!=\!f}(\bm o)
   \! =\!\int_{(\bm x,\bm v)\!\in A(\bm o,f)}\!\!\Pi_{k\!=\!0}^T\Pi_{i\!=\!1}^NP_{O_i^{k}|F\!=\!f,V_i^k\!=\!v_i^k}(o_i^k)d(\bm x,\bm v).
\end{aligned}
$$
Besides, notice for any possible $\bm o $, $f,$ and $f' \in\operatorname{supp}(P_F)$, $P_{O|F=f'}(\bm o)>0$. Then the following equation holds
$$\begin{aligned}
         &\!\!\frac{P_{O|F\!=\!f}(\bm o)}{P_{O|F\!=\!f'\!}(\bm o)}\!\!=\!\!\frac{\int_{(\bm x,\bm v)\!\in\! A(\bm o,f)}\!\!\Pi_{k\!=\!0}^T\Pi_{i\!=\!1}^NP_{O_i^{k}|F\!=\!f,V_i^k\!=\!v_i^k}(o_i^k)d(\!\bm x\!,\!\bm v\!)}{\int_{(\!\bm x'\!,\!\bm v'\!)\!\in \!A(\bm o,f')}\!\!\Pi_{k\!=\!0}^T\Pi_{i\!=\!1}^NP_{O_i^{k}|F\!=\!f',V_i^{k}\!=\!v_i^{\prime k}}(o_i^k)d(\!\bm x'\!,\!\bm v'\!)}\!.
    \end{aligned}$$

    {\color{black}\textbf{Step 2:} We evaluate the likelihood ratio $\frac{P_{O|F=f}(\bm o)}{P_{O|F=f'}(\bm o)}$ under Laplace noise. 
For any $k$, $f$, $f'$, $o^{k}$, $ x^{k}$, and $ v^{k}$,
we have $x_i^{k+1}-x_i^{\prime k+1}=h_x(x_i^{k},v_i^{k},\lambda^{k},f_i)-h_x(x_i^{\prime k},v_i^{\prime k},\lambda^{k},f_i')$, and 
$v_i^{k+1}-v_i^{\prime k+1}=h_v(v_i^k,x_i^{k+1},x_i^k,o^k,\gamma^k)-h_v(v_i^{\prime k},x_i^{\prime k+1},x_i^{\prime k},o^k,\gamma^k)$. 
Therefore,  there exists a unique bijection from  $A(\bm o,f')$ to   $A(\bm o,f)$, }and 
$$    \begin{aligned}
    &\int_{(\bm x,\bm v)\in A(\bm o,f)}\Pi_{k=0}^T\Pi_{i=1}^NP_{O_i^{k}|F=f,V_i^k=v_i^k}(o_i^k)d(\bm x,\bm v)\\
=&\int_{(\bm x', \bm v')\in A(\bm o,f')}\frac{\Pi_{k=0}^T\Pi_{i=1}^NP_{O_i^{k}|F=f,V_i^k=v_i^k}(o_i^k)}{\Pi_{k=0}^T\Pi_{i=1}^NP_{O_i^{k}|F=f',V_i^k=v_i^{\prime k}}(o_i^k)}\\
    &\quad \quad \quad \quad \quad 
    \quad \cdot\Pi_{k=0}^T\Pi_{i=1}^NP_{O_i^{k}|F=f',V_i^k=v_i^{\prime k}}(o_i^k)d (\bm x',\bm v').
\end{aligned}$$
Moreover, since $\zeta_i^k\sim Lap(0,M^k)$, we have
$$
\begin{aligned}
    \frac{P_{O_i^{k}|F=f,V_i^k=v_i^k}(o_i^k)}{P_{O_i^{k}|F=f',V_i^k=v_i^{\prime k}}(o_i^k)}
    =&\frac{\mathbb{P}_{M^k}(o_i^k-v_i^{k})}{\mathbb{P}_{M^k}(o_i^k-v_i^{\prime k})}    = Q_i^k(\bm o,\bm v,\bm v', M^k).
\end{aligned}
$$
Based on the above, we have
    $
\frac{P_{O|F=f}(\bm o)}{P_{O|F=f'}(\bm o)}=\frac{\int_{(\bm x',\bm v')\in A(\bm o,f')}\Pi_{k=0}^T\Pi_{i=1}^NQ_i^k(\bm o,\bm v,\bm v', M^k)P_{O_i^{k}|F=f',V_i^k=v_i^{\prime k}}(o_i^k)d (\bm x',\bm v')}{\int_{(\bm x',\bm v')\in A(\bm o,f')}\Pi_{k=0}^T\Pi_{i=1}^NP_{O_i^{k}|F=f',V_i^{k}=v_i^{\prime k}}(o_i^k)d(\bm x',\bm v')}, $ 
where $(\bm x,\bm v)\in A(\bm o,f)$.$\hfill\square$

\subsection{Proof of Lemma \ref{le::delta}}\label{ap::le::delta}
{\color{black}
For any $\bm o, \bm v, \bm v'$, and $M^k$, notice that$$\begin{aligned}
    -(\|o_i^k-v_i^{k}\|_1-\|o_i^k-v_i^{\prime k}\|_1)\geq& -||o_{i}^k-v_i^{k}-(o_{i}^k-v_i^{\prime k})||_1\\
    =&-\|v_i^{k}-v_i^{\prime k}\|_1,
\end{aligned}$$
and 
$$\begin{aligned}
    -(\|o_i^k-v_i^{k}\|_1-\|o_i^k-v_i^{\prime k}\|_1)\leq& ||o_{i}^k-v_i^{k}-{\color{black}(o_{i}^k-v_i^{\prime k})}||_1\\
    =&\|v_i^{k}-v_i^{\prime k}\|_1.
\end{aligned}$$}
Then
$$\begin{aligned}
\Pi_{k=0}^{T}\Pi_{i=1}^N Q_i^k(\bm o,\bm v,\bm v', M^k)
    \geq&\exp\left(-\sum_{k=0}^T\frac{\sum\limits_{i=1}^N||v_i^{k}-v_i^{\prime k}||_1}{M^k}\right)\\
     \geq& \exp\left(-\sum_{k=0}^T\frac{\Delta(k,f,f',o)}{M^{k}}\right),
\end{aligned}$$
and $\begin{aligned}
\Pi_{k=0}^{T}\Pi_{i=1}^N Q_i^k(\bm o,\bm v,\bm v', M^k)
     \leq& \exp\left(\sum_{k=0}^T\frac{\Delta(k,f,f',o)}{M^{k}}\right).
\end{aligned}$
According to Lemma \ref{le::1}, we have
$\exp\left(-\sum\limits_{k=0}^T\frac{\Delta(k,f,f',\bm o)}{M^{k}}\right)\leq\frac{P_{O|F=f}(\bm o)}{P_{O|F=f'}(\bm o)}\leq \exp\left(\sum\limits_{k=0}^T\frac{\Delta(k,f,f',\bm o)}{M^{k}}\right).$ $\hfill\square$

\subsection{Proof of Theorem \ref{th::determinsistic game upper bound}}\label{ap::th::determinsistic game upper bound}

{\color{black}The proof follows by combining Lemmas \ref{le::delta} and Lemma \ref{le::PMLconvert}.} According to Lemma \ref{le::delta}, for any possible  $ \bm o$, $f,$ and $f'$,  we have
$\frac{P_{O|F=f}(\bm o)}{P_{O|F=f'}(\bm o)}\geq\exp\left(-\sum\limits_{k=0}^T\frac{\Delta(k,f,f',\bm o)}{M^{k}}\right)$.
Then,
 $$\begin{aligned}
    \! \mathbb{E}_{f\sim F}\!\left[\!\frac{P_{O|F=f}(\bm o)}{P_{O|F=f'}(\bm o)}\!\right]
    \!=&\int_{f\in\operatorname{supp}\left(P_F\right)}\frac{P_{O|F=f}(\bm o)}{P_{O|F=f'}(\bm o)}P_F(f)d f\\
    \geq &\!\!\int_{f\!\in\!\operatorname{supp}\!\left(\!P_F\!\right)}\!\!\!\!\exp\!\!\left(\!\!-\!\sum_{k\!=\!0}^T\!\frac{\Delta(k\!,f\!,f'\!,\bm o\!)}{M^{k}}\!\!\right)\!\!P_F(\!f\!)d \!f\\
    =&\mathbb{E}_{f\sim F}\left[\exp(-\sum\limits_{k=0}^T\frac{\Delta(k,f,f',\bm o)}{M^k})\right].
 \end{aligned}$$
Since $\inf\limits_{\scriptsize \substack{\bm o\in\operatorname{supp}\left(\!P_O\!\right) \\ f'\in\operatorname{supp}\left(\!P_F\!\right)}}\mathbb{E}_{f\sim F}\left[\exp(-\sum\limits_{k=0}^T\frac{\Delta(k,f,f',\bm o)}{M^k})\right]\geq \exp(-\epsilon),$ we obtain
$$\inf\limits_{o\in\operatorname{supp}\left(P_O\right)}\inf\limits_{f'\in\operatorname{supp}\left(P_F\right)}\mathbb{E}_{f\sim F}\left[\frac{P_{O|F=f}(\bm o)}{P_{O|F=f'}(\bm o)}\right]\geq \exp(-\epsilon).$$
According to Lemma \ref{le::PMLconvert},
$\begin{aligned}
    \sup\limits_{\bm o\in\operatorname{supp}\left(P_O\right)}\exp \ell(F\to \bm o)\leq \exp(\epsilon).
\end{aligned}$
Therefore, algorithm $\mathcal{A}$ is $\epsilon$-PML. $\hfill\square$

\subsection{Proof of Theorem \ref{th::compare_DP_PML_upper_bound}}\label{ap::th::compare_DP_PML_upper_bound}
 1) Consider that algorithm $\mathcal{A}$ is $\epsilon_1$-DP for groups of size $N$, with $\epsilon_1>0$. 
Then the following result holds:
\begin{equation}\label{eq::DP for groups in proof}
\sup\limits_{\bm o\in\operatorname{supp}\left(P_O\right)}\sup\limits_{f,f'\in\operatorname{supp}\left(P_{F}\right)}\frac{P_{O|F=f}(\bm o)}{P_{O|F=f'}(\bm o)}\leq \exp (\epsilon_1).
\end{equation}
{\color{black}Take the reciprocal of both sides of (\ref{eq::DP for groups in proof}), and we have}
$$\inf_{\bm o\in\operatorname{supp}\left(P_O\right)}\inf_{f,f'\in\operatorname{supp}\left(P_{F}\right)}\frac{P_{O|F=f}(\bm o)}{P_{O|F=f'}(\bm o)}\geq \exp (- \epsilon_1).$$
Since for all $f$,
$\inf\limits_{\bm o\in\operatorname{supp}\left(P_O\right)}\inf\limits_{f'\in\operatorname{supp}\left(P_F\right)}\frac{P_{O|F=f}(\bm o)}{P_{O|F=f'}(\bm o)}\geq \inf\limits_{\bm o\in\operatorname{supp}\left(P_O\right)}\inf\limits_{f,f'\in\operatorname{supp}\left(P_{F}\right)}\frac{P_{O|F=f}(\bm o)}{P_{O|F=f'}(\bm o)}$, by Fatou's lemma,
$$\begin{aligned}
    &\inf\limits_{\bm o\in\operatorname{supp}\left(P_O\right)}\inf\limits_{f'\in\operatorname{supp}\left(P_F\right)}\mathbb{E}_{f\sim F}\left[\frac{P_{O|F=f}(\bm o)}{P_{O|F=f'}(\bm o)}\right]\\
    \geq&\mathbb{E}_{f\sim F}\left[\inf\limits_{\bm o\in\operatorname{supp}\left(P_O\right)}\inf\limits_{f'\in\operatorname{supp}\left(P_F\right)}\frac{P_{O|F=f}(\bm o)}{P_{O|F=f'}(\bm o)}\right]\\
    \geq &\exp (- \epsilon_1).
\end{aligned}$$ 
{\color{black}According to Lemma \ref{le::PMLconvert},  algorithm $\mathcal{A}$ is $\epsilon_2$-PML and $\epsilon_2\leq\epsilon_1$.}

2) Consider that  $\sum\limits_{k=0}^T\frac{\Delta''(k)}{M^k}\leq \epsilon_1,$
According to Lemma \ref{le::delta}, 
$\begin{aligned}
   \sup\limits_{\scriptsize \substack{\bm o\in\operatorname{supp}\left(P_O\right)\\
   f,f'\in\operatorname{supp}\left(P_{F}\right)}}\frac{P_{O|F=f}(\bm o)}{P_{O|F=f'}(\bm o)}
    \leq &\exp\left(\sum\limits_{k=0}^T\frac{\Delta''(k)}{M^k}\right).    
\end{aligned}$
Then the algorithm is $\epsilon_1$-DP for groups of size $N$. Similarly,
$$\begin{aligned}
    &\inf\limits_{\scriptsize \substack{\bm o\in\operatorname{supp}\left(\!P_O\!\right) \\ f'\in\operatorname{supp}\left(\!P_F\!\right)}}\mathbb{E}_{f\sim F}\left[\exp(-\sum\limits_{k=0}^T\frac{\Delta(k,f,f',\bm o)}{M^k})\right]\\
    \geq &\mathbb{E}_{f\sim F}\left[\exp(-\sum\limits_{k=0}^T\frac{\sup\limits_{\scriptsize \substack{\bm o\in\operatorname{supp}\left(\!P_O\!\right) \\ f'\in\operatorname{supp}\left(\!P_F\!\right)}}\Delta(k,f,f',\bm o)}{M^k})\right]\\
    \geq &\exp (- \epsilon_1).
\end{aligned}$$ 
{\color{black}According to Theorem \ref{th::determinsistic game upper bound},  algorithm $\mathcal{A}$ is $\epsilon_2$-PML and $\epsilon_2\leq\epsilon_1$.}

\subsection{Proof of Lemma \ref{le::eq_different}}\label{ap::le::eq_different}
According to Lemma \ref{le::1}, for any $i\in[N]$,$f_i\neq f_i'$,$f_{-i}$,
 $$\begin{aligned}
&\frac{P_{O|F=\{f_i,f_{-i}\}}(\bm o)}{P_{O|F=\{f_i',f_{-i}\}}(\bm o)}\\
         =&\!\frac{\!\!\int_{\!(\!\bm x'\!,\bm v'\!)\!\in\! A(\!\bm o\!,\!\{f_i'\!,f_{\!-\!i}\!\}\!)}\! \!\Pi_{k\!=\!0}^T\!\Pi_{j\!=\!1}^NQ_j^k\!(\!\bm o\!,\!\bm v\!,\!\bm v'\!\!, \!M^k\!)\!P_{\!\!O^{k}\!|\!F\!=\!f'\!\!,\!V_j^k\!=\!v_j^{\prime k}}(\!o_j^k\!)d (\!\bm x'\!\!,\!\bm v'\!)}{\int_{(\bm x'\!,\!\bm v')\!\in\! A(\bm o\!,\!\{f_i',f_{\!-\!i}\})}\Pi_{k\!=\!0}^T\Pi_{j\!=\!1}^NP_{O_j^{k}|F\!=\!f'\!,\!V_i^{k}\!=\!v_j^{\prime k}}(o_j^k)d(\bm x'\!,\bm v')}\!.       
    \end{aligned}$$
For $j\neq i$, note $x_j^{0}=x_j^{\prime 0}$, and $v_j^{0}=v_j^{\prime 0}$ in algorithm $\mathcal{A}$. If $x_j^{k}=x_j^{\prime k}$ and $v_j^{k}=v_j^{\prime k}$, then $x_j^{k+1}-x_j^{\prime k+1}=h_x(x_j^{k},v_j^{k},\lambda^{k},f_j)-h_x(x_j^{\prime k},v_j^{\prime k},\lambda^{k},f_j)\!=\!0$, and 
$v_j^{k\!+\!1}\!-\!v_j^{\prime k\!+\!1}\!=\!h_v(v_j^k,x_j^{k+1},x_j^{k},o^k,\gamma^k)-h_v(v_j^{\prime k},x_j^{\prime k+1},x_j^{\prime k},o^k,\gamma^k)=0$. Thus, $x_j^{k}=x_j^{\prime k}$ and $v_j^{k}=v_j^{\prime k}$ for all $k$. Then $Q_j^k(\bm o,\bm v,\bm v', M^k)=\exp(-\frac{||o_{j}^k-v_i^{k}||_1-||o_{j}^k-v_j^{\prime k}||_1}{M^k})=1$. Therefore,  $\frac{P_{O|F_i=f_i,F_{-i}=f_{-i}}(\bm o)}{P_{O|F_i=f_i',F_{-i}=f_{-i}}(\bm o)}
         =\frac{\int_{(\bm x'\!,\bm v')\in A(\bm o\!,\!\{f_i'\!,f_{\!-\!i}\!\}\!)}\! \Pi_{k\!=\!0}^TQ_i^k(\bm o,\bm v,\bm v', M^k)\Pi_{j\!=\!1}^NP_{O_j^{k}|F\!=\!f'\!,V_j^k\!=\!v_j^{\prime k}}(\!o_j^k\!)d \!(\!\bm x'\!,\bm v'\!)}{\int_{(\bm x',\bm v')\in A(\bm o,\{f_i',f_{-i}\})}\Pi_{k=0}^T\Pi_{j=1}^NP_{O_j^{k}|F=f',V_j^{k}=v_j^{\prime k}}(o_j^k)d(\bm x',\bm v')}\!,    $
         where $(\bm x,\bm v)\in A(\bm o,f)$. $\hfill\square$


\subsection{Proof of Lemma \ref{th::determinsistic game lower bound}}\label{ap::th::determinsistic game lower bound}

(1) {\color{black}We first derive a bound on $\frac{P_{O|F=f}(\bm o)}{P_{O|F=f'}(\bm o)}$ for any $f$ and $f'$. We then use this result to establish a lower bound on $\frac{P_{O|F_1=f^1}(\bm o)}{P_{O|F_1=f^0}(\bm o)}$, which yields an upper bound on $\exp(\ell(F_1\to \bm o))$.}

{\color{black}\textbf{Step 1: } We present the bound on $\frac{P_{O|F=f}(\bm o)}{P_{O|F=f'}(\bm o)}$ with $\epsilon$. By Definition \ref{def::DP}, for any $\bm o$, $i\in[N]$, $f_i,f_i'$, and $f_{-i}$ in support sets, 
$\frac{P_{O|F=\{f_i,f_{-i}\}}(\bm o)}{P_{O|F=\{f_i',f_{-i}\}}(\bm o)}\leq \exp (\epsilon)$.
For any $f=\{f_1,\dots,f_N\}$, take the set $f_{[i:j]}=\{f_i,\dots,f_j\}$, if $i\leq j$. Otherwise, take $f_{[i:j]}=\varnothing$. For any $f$ and $f'$, we have
$
    \begin{aligned}
        \frac{P_{O|F=f}(\bm o)}{P_{O|F=f'}(\bm o)}
        =&\Pi_{i=1}^{N}\frac{P_{O|F=\{f_{[1:i-1]}',f_i,f_{[i+1:N]}\}}(\bm o)}{P_{O|F=\{{f_{[1:i-1]}',f_i',f_{[i+1:N]}\}}}(\bm o)}.
    \end{aligned}
$}
Note $\exp(-\epsilon)\leq\frac{P_{O|F=\{f_{[1:i-1]}',f_i,f_{[i+1:N]}\}}(\bm o)}{P_{O|F=\{{f_{[1:i-1]}',f_i',f_{[i+1:N]}\}}}(\bm o)}\leq\exp(\epsilon)$. If $f_i= f_i'$, then $\frac{P_{O|F=\{f_{[1:i-1]}',f_i,f_{[i+1:N]}\}}(\bm o)}{P_{O|F=\{{f_{[1:i-1]}',f_i',f_{[i+1:N]}\}}}(\bm o)}=1$. Define the Hamming distance between $f$ and $f'$, i.e., 
   $ D_H(f,f')=\sum\limits_{i=1}^N \mathbf{1}_{f_i\neq f_{i}'}.$
Thus, $ \exp (-D_H(f,f') \epsilon)\leq\frac{P_{O|F=f}(\bm o)}{P_{O|F=f'}(\bm o)}\leq \exp (D_H(f,f') \epsilon).$
{\color{black}Since the number of $f_{-1}
\in \mathcal{F}_{-1}$ such that $\sum\limits_{i=2}^N \mathbf{1}_{f_i\neq f^0}=j$ is $\binom{N-1}{j}$, }
\begin{equation}\label{eq::ap::th::3::01}
    \begin{aligned}&\!\sum\limits_{f_{\!-\!1}\in\mathcal{F}_{\!-\!i}}\!\frac{P_{O|F_1=f^1\!,F_{\!-\!1}=f_{\!-\!1}}(\!\bm o\!)}{P_{O|F=(f^0)^{N}}(\bm o)}\!\geq\!\!\sum\limits_{j=0}^{N-1}\!\!\binom{N\!-\!1}{j}\!\!\exp(\!-\!\epsilon)^{1\!+\!j\!},\!\end{aligned}
\end{equation}
\begin{equation}\label{eq::ap::th::3::02}
    \begin{aligned}
&\sum\limits_{f_{\!-\!1}\in\mathcal{F}_{\!-\!i}}\frac{P_{O|F_1=f^0,F_{\!-\!1}=f_{\!-\!1}}(\bm o)}{P_{O|F=(f^0)^{N}}(\bm o)}\!\leq\!\sum\limits_{j=0}^{N\!-\!1}\!\binom{N\!-\!1}{j}\!\exp(\epsilon)^j.\!\end{aligned}
\end{equation}

{\color{black}\textbf{Step 2:} We use the above result to establish a lower bound on $\frac{P_{O|F_1=f^1}(\bm o)}{P_{O|F_1=f^0}(\bm o)}$. }By expanding $P_{O|F_1=f^1}(\bm o)$ using (\ref{eq::th3::P_F{-1}}), we have
\begin{equation}\label{eq::ap::th::3::111}
    \begin{aligned}
    &\frac{P_{O|F_1=f^1}(\bm o)}{P_{O|F=(f^0)^{N}}(\bm o)}\\
    =&\!\beta\frac{ P_{O|F\!=\!(\!f^1\!)^{\!N}}\!(\!\bm o)}{P_{O|F\!=\!(\!f^0\!)^{\!N}}\!(\!\bm o)}\!\!+\!\!\frac{1\!-\!\beta}{2^{N\!\!-\!\!1}\!\!-\!\!1}\!\!\!\left(\!\!\sum\limits_{f_{\!-\!1}\in\mathcal{F}_{\!-\!i}}\!\!\!\!\!\frac{ P_{O|F\!=\!\{f^1\!\!,f_{\!-\!1}\}}\!(\!\bm o)}{P_{O|F=(\!f^0\!)^{N}}\!(\!\bm o)}\!\!-\!\!\frac{ P_{O|F\!=\!(\!f^1\!)^N}\!(\!\bm o)}{P_{O|F\!=\!(\!f^0\!)^{N}}\!(\!\bm o)}\!\!\!\right)\\
    =&\!\!\left(\!\!\beta\!-\!\frac{1\!-\!\beta}{2^{\!N\!-\!1}\!-\!1}\!\!\right)\!\!\frac{P_{\!O|F\!=\!(f^{\!1})^{\!N}}(\!\bm o)}{P_{\!O|F\!=\!(f^{\!0})^{\!N}}(\!\bm o)}\!+\!\frac{1\!-\!\beta}{2^{\!N\!-\!1}\!\!-\!1}\!\!\!\sum\limits_{f_{\!-\!1}\in\mathcal{F}_{\!-\!i}}\!\!\!\!\!\frac{P_{\!O|F_1\!=\!f^{\!1}\!,F_{\!-\!1}\!=\!f_{\!-\!1}}(\!\bm o)}{P_{O|F\!=\!(f^0)^{N}}(\!\bm o)}\!.
\end{aligned}
\end{equation}

Combining (\ref{eq::ap::th::3::01}) and (\ref{eq::ap::th::3::111}), we obtain
    \begin{equation}\label{eq::111}\begin{aligned}
    &\frac{P_{O|F_1=f^1}(\bm o)}{P_{O|F=(f^0)^{N}}(\bm o)}\\
    \geq&\left(\!\beta\!-\!\frac{1\!-\!\beta}{2^{N\!-\!1}\!\!-\!1}\!\right)\!\exp(\!-\!\epsilon)^N\!\!+\!\frac{1\!-\!\beta}{2^{N\!-\!1}\!\!-\!1}\sum\limits_{j=0}^{N\!-\!1}\!\binom{N\!-\!1}{j}\exp\!(\!-\!\epsilon)^{1\!+\!j}\\
    =& \!\exp(\!-\!\epsilon)\!\!\!\left(\!\!\!\left(\!\!\beta\!-\!\frac{1\!-\!\beta}{2^{\!N\!-\!1}\!-\!1}\!\!\right)\!\!\exp(\!-\!\epsilon\!)^{\!N\!-\!1}\!\!+\!\!\!\frac{1\!-\!\beta}{2^{\!N\!-\!1\!}\!\!-\!1}\!\!\sum\limits_{j\!=\!0}^{\!N\!-\!1}\!\!\binom{\!\!N\!-\!1\!\!}{j}\!\exp(\!\!-\!\epsilon\!)^{\!j}\!\!\!\right)\\
    =&\!\exp(\!-\!\epsilon)\!\!\left(\!\!\!\left(\!\beta\!-\!\!\frac{1\!-\!\beta}{2^{N\!-\!1}\!\!-\!1}\right)\!\!\exp(\!-\!\epsilon)^{\!N\!-\!1}\!\!
  \!+\!\frac{1\!-\!\beta}{2^{N\!-\!1}\!\!-\!\!1}\!\!\left(1\!+\!\exp(\!-\!\epsilon)\!\right)^{\!N\!-\!1}\!\!\!\right)\!.
\end{aligned}
\end{equation}

Similarly to (\ref{eq::ap::th::3::111}), we have
\begin{equation}\label{eq::ap::th::3::222}
    \begin{aligned}
    \frac{P_{O|F_1\!=\!f^0}(\bm o)}{P_{O|F\!=\!(f^0)^{N}}(\bm o)}\!
    =&\left(\beta-\frac{1-\beta}{2^{N-1}-1}\right)\frac{P_{O|F=(f^0)^{N}}(\bm o)}{P_{O|F=(f^0)^{N}}(\bm o)}\\
    &+\!\frac{1\!-\!\beta}{2^{N\!-\!1}\!-\!1}\!\!\!\sum\limits_{f_{\!-1}\in\mathcal{F}_{-i}}\!\!\!\!\!\frac{P_{O|F_1\!=\!f^0\!,F_{\!-\!1}\!=\!f_{\!-\!1}}\!(\bm o)}{P_{O|F=(f^0)^{N}}(\bm o)}.\\
\end{aligned}
\end{equation}
Based on (\ref{eq::ap::th::3::02}) and (\ref{eq::ap::th::3::222}), we have the following inequality:
$$\begin{aligned}
    &\frac{P_{O|F_1=f^0}(\bm o)}{P_{O|F=(f^0)^{N}}(\bm o)}\\
    \leq&
     \left(\beta-\frac{1-\beta}{2^{N-1}-1}\right)+\frac{1-\beta}{2^{N-1}-1}\sum\limits_{j=0}^{N}\binom{N-1}{j}\exp(\epsilon)^j\\
    =&\left(\beta-\frac{1-\beta}{2^{N-1}-1}\right)+\frac{1-\beta}{2^{N-1}-1}\left(1+\exp(\epsilon\right))^{N-1}.
\end{aligned}$$
Recalling that $\frac{P_{O|F_1=f^1}(\bm o)}{P_{O|F_1=f^0}(\bm o)}
    =\frac{\frac{P_{O|F_1=f^1}(\bm o)}{P_{O|F=(f^0)^{N}}(\bm o)}}{\frac{P_{O|F_1=f^0}(\bm o)}{P_{O|F=(f^0)^{N}}(\bm o)}}$, we get
\begin{equation}\label{eq::th3:1231232}
    \begin{aligned}
   & \frac{P_{O|F_1=f^1}(\bm o)}{P_{O|F_1=f^0}(\bm o)}\\
    \geq&\!\frac{\!\exp(\!-\!\epsilon)\!\left(\!\left(\!\beta\!-\!\frac{1\!-\!\beta}{2^{N\!-\!1}\!-\!1}\!\right)\!\exp(\!-\!\epsilon\!)^{N\!-\!1}\!+\!\frac{1\!-\!\beta}{2^{N\!-\!1}\!-\!1}\!\left(1\!+\!\exp(\!-\!\epsilon\!\right)^{N\!-\!1}\!\right)}{\left(\beta-\frac{1-\beta}{2^{N-1}-1}\right)+\frac{1-\beta}{2^{N-1}-1}\left(1+\exp(\epsilon)\right)^{N-1}}\\
    =&\!\frac{\!\exp(\!-\!\epsilon\!)\!\left(\!\left(2^{N\!-\!1}\!-\!1\!\right)\!\exp(\!-\!\epsilon\!)^{N\!-\!1}\!+\!(1\!-\!\beta)\!\left(\!1\!+\!\exp(\!-\!\epsilon\!)\!\right)^{N\!-\!1}\!\right)}{\exp(\!\epsilon\!)^{N\!-\!1}\left(\!\left(\!2^{N\!-\!1}\!-\!1\right)\exp(\!-\!\epsilon\!)^{N\!-\!1}\!+\!(1\!-\!\beta)\left(1\!+\!\exp(\!-\!\epsilon)\!\right)^{N\!-\!1}\!\right)}\\
    =& \exp(-N\epsilon).
\end{aligned}
\end{equation}
By exchanging $f^1$ and $f^0$ in the above, we get $\frac{P_{O|F_1=f^0}(\bm o)}{P_{O|F_1=f^1}(\bm o)}\geq\exp(-N\epsilon)$.  Recalling Proposition \ref{Pro::1} and $\alpha<0.5$, we have
$$ 
\begin{aligned}
    \exp(\ell(F_1\to \bm o))
    \leq &\frac{1}{\min\limits_{f_1\in\mathcal{F}_1}\frac{(1-\alpha)P_{O|F_1=f^1}(\bm o)+\alpha P_{O|F_1=f^0}(\bm o)}{P_{O|F_1=f_1}(\bm o)}},
\end{aligned}$$
and
$
   \sup\limits_{\bm o\in\operatorname{supp}\left(P_O\right)}\exp(\ell(F_1\to \bm o))\leq\frac{1}{(1\!-\!\alpha)\exp(-N\epsilon)+\!\alpha}.
$

(2) {\color{black}In the following, we first construct an observation sequence $\bm o$ satisfying $\exp(\epsilon_o)>1$, and then exploit the exact values of certain  ratios $\frac{P_{O|F=f}(\bm o)}{P_{O|F=f'}(\bm o)}$. Based on these results, we derive a closed-form expression for $\frac{P_{O|F_1=f^1}(\bm o)}{P_{O|F_1=f^0}(\bm o)}$ and substitute it into the PML expression to obtain the  lower bound.}

{\color{black}\textbf{Step 1:} We construct an observation
sequence $\bm o \in\operatorname{supp}\left(P_O\right)$, where $o^k_i=o^k_j$ for $i,j\in[N]$.} Take 
$
    \exp(\epsilon_o)\!=\!
    \max_{i\in[N]}\!\sup_{\scriptsize \substack{f_{\!-\!i}\in \!\operatorname{supp}\left(P_{F_{\!-\!i}}\right) \\ f_i,\!f_i'\in \!\operatorname{supp}\left(P_{F_i}\right)}}\!\!\frac{P_{O|F_i\!=\!f_i,F_{\!-\!i}=f_{\!-\!i}}(\bm o)}{P_{O|F_i\!=\!f_i',F_{\!-\!i}\!=\!f_{\!-i\!}}(\bm o)}.
$
Note that there exists  $\bar{\bm o}=\left\{\{\bar{o}_i^k\}_{i=1}^N\right\}_{k=0}^T\in\operatorname{supp}\!\left(\!P_O\!\right)$ such that
$\!\max\limits_{i\in[N]}\!\sup_{\scriptsize \substack{f_{\!-\!i}\in \!\operatorname{supp}\left(P_{F_{\!-\!i}}\right) \\ f_i,\!f_i'\!\in \!\operatorname{supp}\left(P_{F_i}\right)}}\frac{P_{O|F_i\!=\!f_i,F_{\!-\!i}=f_{\!-\!i}}(\bar{\bm o})}{P_{O|F_i\!=\!f_i',F_{\!-\!i}\!=\!f_{\!-i\!}}(\bar{\bm o})}>1.$
Since $F_i=\{f^0,f^1\}$ for all $i\in [N]$, the previous inequality can be obtained for $i=1$. 
 Take another observation sequence $\bm o =\left\{\{o_i^k\}_{i=1}^N\right\}_{k=0}^T\in\!\operatorname{supp}\!\left(\!P_O\!\right)$ where $o_i^k=\bar{o}_1^k$ for all $i\in[N]$. Recalling Lemma \ref{le::eq_different}, we have $\frac{P_{O|F_1\!=\!f_1,F_{\!-\!1}=f_{\!-\!1}}(\bar{\bm o})}{P_{O|F_1\!=\!f_1',F_{\!-\!1}\!=\!f_{\!-1\!}}(\bar{\bm o})}=\frac{P_{O|F_1\!=\!f_1,F_{\!-\!1}=f_{\!-\!1}}({\bm o})}{P_{O|F_1\!=\!f_1',F_{\!-\!1}\!=\!f_{\!-1\!}}({\bm o})}.$ Thus, $\exp(\epsilon_o)>1$.



{\color{black}\textbf{Step 2:} We exploit the exact
values of  certain  ratios $\frac{P_{O|F=f}(\bm o)}{P_{O|F=f'}(\bm o)}$   by  $\exp(-\epsilon_o)$.}
Notice that 
$\sup_{f_{-1}\in \operatorname{supp}\left(P_
{F_{-1}}\right)}\frac{P_{O|F_1=f^0,F_{-1}=f_{-1}}(\bm o)}{P_{O|F_i=f^1,F_{-1}=f_{-1}}(\bm o)}=\exp(\epsilon_o).$
According to Lemma \ref{le::eq_different}, for any $f_{-1}$, we have $\frac{P_{O|F_1=f^0,F_{-1}=f_{-1}}(\bm o)}{P_{O|F_i=f^1,F_{-1}=f_{-1}}(\bm o)}=\exp(\epsilon_o).$ Then for any $f_{-1}$,
$\exp(-\epsilon_o)=\frac{P_{O|F_1=f^1,F_{-1}=f_{-1}}(\bm o)}{P_{O|F_i=f^0,F_{-1}=f_{-1}}(\bm o)}<1.$ According to Lemma \ref{le::1} and Lemma \ref{le::eq_different}, the values of $\exp(-\epsilon_o)$ and $\frac{P_{O|F_1=f^1,F_{-1}=f_{-1}}(\bm o)}{P_{O|F_1=f^0,F_{-1}=(f^0)^{N-1}}(\bm o)} $ are influenced by $Q_i^k\!(\!\bm o\!,\!\bm v\!,\!\bm v'\!\!, \!M^k\!)$. 
Actually, for any $i\neq 1$, similar to the proof of Lemma \ref{le::eq_different}, we have $Q_i^k\!(\!\bm o\!,\!\bm v\!,\!\bm v'\!\!, \!M^k\!)=1$ if $f_i =f^0$. Additionally, if $f_i =f^1$, similar to the proof in Lemma \ref{le::eq_different}, recalling  $x_i^{0}=x_1^{0}$ and $v_i^{0}=v_1^{0}$,
we have $x_i^{k}=x_1^{k}$ and $v_i^{k}=v_1^{k}$ for all $k$. Also, we can get $x_i^{\prime k}=x_1^{\prime k}$ and $v_i^{\prime k}=v_1^{\prime k}$. 
Then $Q_i^k(\bm o,\bm v,\bm v', M^k)=\exp(-\frac{||o_{i}^k-v_i^{k}||_1-||o_{i}^k-v_i^{\prime k}||_1}{M^k})= \exp(-\frac{||o_{1}^k\!-\!v_1^{k}||_1\!-\!||o_{1}^k\!-\!v_1^{\prime k}||_1}{M^k})\!=\!Q_1^k(\bm o,\bm v,\bm v', M^k)$,
and $\Pi_{i=1}^N\!Q_i^k(\bm o,\bm v,\bm v', M^k)=Q_1^k(\bm o,\bm v,\bm v', M^k)^{1+{D_H\left(f_{-1},(f^0)^{N-1}\right)}}$. Thus,
\begin{equation}\label{eq::f^1}
    \begin{aligned}
    \frac{P_{O|F_1\!=\!f^1\!,F_{\!-\!1}\!=\!f_{\!-\!1}}(\bm o)}{P_{O|F_1\!=\!f^0\!,F_{\!-\!1}\!=\!(f^0)^{N\!-\!1}}(\bm o)}
\!=\!\exp\!\left(\!-\!\epsilon_o\!\right)^{1\!+\!{D_H\left(f_{\!-\!1}\!,(f^0)^{N\!-\!1}\!\right)}}.
\end{aligned}
\end{equation}
With the same process, we obtain the following equation \begin{equation}\label{eq::f^0}
    \begin{aligned}
&\frac{P_{O|F_1\!=\!f^0\!,F_{\!-\!1}=f_{\!-\!1}}(\!\bm o)}{P_{O|F_1\!=\!f^0,F_{\!-\!1}=(f^0)^{N\!-\!1}}\!(\!\bm o)}\!=\!\exp\!\left(\!-\!\epsilon_o\!\right)^{\!D_H\left(\!f_{-1},(f^0)^{\!N\!-1\!}\right)}\!.
\end{aligned}
\end{equation}


{\color{black}\textbf{Step 3: } We derive a closed-form expression for $\frac{P_{O|F_1=f^1}(\bm o)}{P_{O|F_1=f^0}(\bm o)}$. }
Since $ P_{F_1}(f^0)=1-P_{F_1}(f^1)=\alpha$, 
\begin{equation}\label{eq::lam5:finalF_1}
    \begin{aligned}
\exp(\ell(F_1\!\to\! \bm o))\!
    \geq\! &\frac{P_{O|F_1=f^0}(\bm o)}{(1\!-\!\alpha)P_{O|F_1=f^1}(\bm o)\!+\!\alpha P_{O|F_1=f^0}(\bm o)}\\
    =&\frac{1}{(1-\alpha)\frac{P_{O|F_1=f^1}(\bm o)}{P_{O|F_1=f^0}(\bm o)}+\alpha }.
\end{aligned}
\end{equation}
Based on (\ref{eq::f^1}), we have
$\frac{P_{O|F=(f^1)^N}(\bm o)}{P_{O|F=(f^0)^{N}}(\bm o)}=\exp(-N\epsilon_o).$
Since the number of ${D_H\left(f_{-1},(f^0)^{N-1}\right)}=j$ for all $f_{-1}\in \{f^0,f^1\}^{N-1}$ is $\binom{N-1}{j}$ and  
$$\begin{aligned}&\sum\limits_{f_{-1}\in\mathcal{F}_{-1}}\frac{P_{O|F_1=f^1,F_{-1}=f_{-1}}(\bm o)}{P_{O|F=(f^0)^{N}}(\bm o)}=\sum\limits_{j=0}^{N-1}\binom{N-1}{j}\exp(-\epsilon_o)^{1+j}.\end{aligned}$$
Besides, recalling the process in (\ref{eq::ap::th::3::111}), we have 
$$\begin{aligned}
    &\frac{P_{O|F_1=f^1}(\bm o)}{P_{O|F=(f^0)^{N}}(\bm o)}\\
    =&\! \!\exp(\!-\!\epsilon_o\!)\!\!\!\left(\!\!\!\left(\!\!\beta\!-\!\frac{1\!-\!\beta}{2^{\!N\!-\!1}\!-\!\!1}\!\!\right)\!\!\exp(\!-\!\epsilon_o\!)^{\!N\!-\!1}\!\!+\!\frac{1\!-\!\beta}{2^{\!N\!-\!1}\!-\!1}\!\!\sum\limits_{j\!=\!0}^{N\!-\!1}\!\!\binom{\!N\!-\!1\!}{j}\!\!\exp(\!-\!\epsilon_o\!)^j\!\!\!\right)\\
    =&\!\!\exp(\!-\!\epsilon_o\!)\!\left(\!\!\!\left(\!\beta\!-\!\frac{1\!-\!\beta}{2^{N\!-\!1}\!-\!\!1}\!\right)\!\exp(\!-\!\epsilon_o)^{\!N\!-\!1}\!+\!\frac{1\!-\!\beta}{2^{\!N\!-\!\!1}\!-\!1}\!\left(\!1\!+\!\exp(\!-\!\epsilon_o\!)\!\right)^{N\!-\!1}\!\!\right)\!.
\end{aligned}$$
Based on (\ref{eq::ap::th::3::222}) and (\ref{eq::f^0}), we obtain
$$\begin{aligned}
    \frac{P_{O|F_1\!=\!f^0}(\bm o)}{P_{O|F\!=\!(f^0)^{\!N}}(\!\bm o)}
    =&\left(\beta\!-\!\frac{1\!-\!\beta}{2^{N\!-\!1}\!-\!1}\right)\!+\!\frac{1\!-\!\beta}{2^{N\!-\!1}\!-\!1}\left(1\!+\!\exp(\!-\!\epsilon_o)\right)^{N\!-\!1}.
\end{aligned}$$
Therefore, 
similar to (\ref{eq::th3:1231232}), we can obtain
$$\begin{aligned}
   & \frac{P_{O|F_1=f^1}(\bm o)}{P_{O|F_1=f^0}(\bm o)}\\
    =&\!\frac{\!\exp(\!-\!\epsilon_o)\!\left(\!\left(\!\beta\!-\!\frac{1\!-\!\beta}{2^{N\!-\!1}\!-\!1}\!\right)\!\exp(\!-\!\epsilon_o\!)^{N\!-\!1}\!+\!\frac{1\!-\!\beta}{2^{N\!-\!1}\!-\!1}\!\left(1\!+\!\exp(\!-\!\epsilon_o\!)\!\right)^{N\!-\!1}\!\right)}{\left(\beta-\frac{1-\beta}{2^{N-1}-1}\right)+\frac{1-\beta}{2^{N-1}-1}\left(1+\exp(-\epsilon_o)\right)^{N-1}}\\
    =&\!\frac{\!\exp(\!-\!\epsilon_o\!)\!\left(\!\left(2^{N\!-\!1}\!-\!1\!\right)\!\exp(\!-\!\epsilon_o\!)^{N\!-\!1}\!+\!(1\!-\!\beta)\!\left(\!1\!+\!\exp(\!-\!\epsilon_o\!)\!\right)^{N\!-\!1}\!\right)}{\left(\!2^{N\!-\!1}\!-\!1\right)\!+\!(1\!-\!\beta)\left(1\!+\!\exp(\!-\!\epsilon_o)\right)^{N-1}}.
\end{aligned}$$
Recalling (\ref{eq::lam5:finalF_1}), we get that  a lower bound of $\exp(\ell(F_1\to \bm o))$ is $\frac{1}{(1\!-\!\alpha)\frac{\exp(\!-\!\epsilon_o)\left(\!\left(2^{N\!-\!1}\!-\!1\!\right)\exp(\!-\!\epsilon_o)^{N\!-\!1}\!+\!(1\!-\!\beta)\left(\!1\!+\!\exp(\!-\!\epsilon_o)\!\right)^{N\!-\!1}\!\right)}{\left(2^{N\!-\!1}\!-\!1\right)\!+\!(1\!-\!\beta)\left(\!1\!+\!\exp(\!-\!\epsilon_o)\!\right)^{N\!-\!1}}\!+\!\alpha}.$

(3) Noticing that $0<1-\beta(N)<1$, we obtain
$$\begin{aligned}
    &\frac{\left(\!\left(2^{N\!-\!1}\!-\!1\!\right)\exp(\!-\!\epsilon_o)^{N\!-\!1}\!+\!(1\!-\!\beta(N))\left(\!1\!+\!\exp(\!-\!\epsilon_o)\!\right)^{N\!-\!1}\!\right)}{\left(2^{N\!-\!1}\!-\!1\right)\!+\!(1\!-\!\beta(N))\left(\!1\!+\!\exp(\!-\!\epsilon_o)\!\right)^{N\!-\!1}}\\
    < &\frac{\!\left(2^{N\!-\!1}\!-\!1\!\right)\exp(\!-\!\epsilon_o)^{N\!-\!1}\!+\left(\!1\!+\!\exp(\!-\!\epsilon_o)\!\right)^{N\!-\!1}}{2^{N\!-\!1}\!-\!1}\\
    =& \exp(\!-\!\epsilon_o)^{N\!-\!1}+\frac{\left(\frac{\!1\!+\!\exp(\!-\!\epsilon_o)}{2}\!\right)^{N\!-\!1}}{1-\!\frac{1}{2^{N-1}}}.
\end{aligned}$$
Noticing that $\exp(-\epsilon_0)<1$ and $0<\frac{1+\exp(-\epsilon_o)}{2}<1$,  we have $\lim_{N\to \infty}\exp(\!-\!\epsilon_o)^{N\!-\!1}+\frac{\left(\frac{\!1\!+\!\exp(\!-\!\epsilon_o)}{2}\!\right)^{N\!-\!1}}{1-\!\frac{1}{2^{N-1}}}=0.$ Then 
$$
    \lim_{N\!\to \!\infty}\!\!\!\!\frac{\!\!\left(2^{N\!-\!1}\!-\!1\!\right)\exp(\!-\!\epsilon_o)^{N\!-\!1}\!+\!(1\!-\!\beta(N))\left(\!1\!+\!\exp(\!-\!\epsilon_o)\!\right)^{N\!-\!1}}{\left(2^{N\!-\!1}\!-\!1\right)\!+\!(1\!-\!\beta(N))\left(\!1\!+\!\exp(\!-\!\epsilon_o)\!\right)^{N\!-\!1}}
    =0.$$
Recalling that $\exp(\ell(F_1\to \bm o))\leq \exp(\epsilon_{\max }\left(F_1\right))=\frac{1}{\alpha}$, we have 
$\lim_{N\to \infty}\exp(\ell(F_1\to \bm o))=\frac{1}{\alpha}.$ $\hfill\square$

\subsection{Proof of Theorem \ref{th::compare_DP_PML_lower_boundd}}\label{ap::th::compare_DP_PML_lower_boundd}

{\color{black}
We aim to show that there exist binary cost function sets and parameters such that the algorithm is $\epsilon_1$-DP, and then prove that applying the correlation in Lemma \ref{th::determinsistic game lower bound} to the cost function sets yields $\epsilon_2$ with PML guarantees.}
For any player's number $N$, consider the same binary database in Lemma \ref{th::determinsistic game lower bound}. We consider a special algorithm $\mathcal{A}$ following the approach in \cite{ye2021differentially}. The strategy update for the $i$th player at iteration $k$ is given by $x_i^{k+1}=x_i^k-\lambda^k F_i(x_i^k,v_i^k)$, while the estimated aggregated term is updated as $v_i^{k+1}=\sum\limits_{j=1}^N\frac{1}{N}o_j^k +x_i^{k+1}-x_i^k.$
According to \cite[Theorem 3]{ye2021differentially}, if $|\nabla_{x_i} \bar{f}_{i}(x_i,\delta(x))|\leq C$, $\lambda^k=cq^k$, $M^k=d\bar{q}^k$, $\bar{q}\in (q,1)$,  then  algorithm $\mathcal{A}$ is $\frac{2cC\bar{q}}{d(\bar{q}-q)}$-DP. Take $f^0,f^1$ such that $C=\frac{d(\bar{q}-q)}{2c\bar{q}}\epsilon_1$. Then the algorithm is $\epsilon_1$-DP for any $N$.
Moreover, according to Lemma \ref{th::determinsistic game lower bound}, there exists $ \bm o  \in \operatorname{supp}\left(P_O\right)$ such that 
$\lim_{N\to \infty}\ell(F_1\to \bm o)=\log\left(\frac{1}{\alpha}\right).$
Notice that in this case, $\max\limits_{i\in[N]}\epsilon_{\max }(F_i)=\log(\frac{1}{\alpha})$. Then
$\lim_{N\to \infty}\max\limits_{i\in[N]}\ell (F_i\to \bm o)=\max\limits_{i\in[N]}\epsilon_{\max }(F_i).$
For any $\epsilon_2\in(0,\max\limits_{i\in[N]}\epsilon_{\max }(F_i))$, there exists $N$ such that
$\max\limits_{i\in[N]}\ell (F_i\to \bm o)\geq\epsilon_2.$ 
Thus, for $\epsilon_1$ and $\epsilon_2\in(0,\max\limits_{i\in[N]}\epsilon_{\max }(F_i))$, we can construct  correlated datasets in Lemma \ref{th::determinsistic game lower bound}, such that  algorithm $\mathcal{A}$ is $\epsilon_1$-DP, and $\sup\limits_{\bm o  \in \operatorname{supp}\left(P_O\right)}\max\limits_{i\in[N]}\ell (F_i\to \bm o)\geq\epsilon_2.$
$\hfill\square$

\section*{References}
\bibliographystyle{IEEEtran}
\bibliography{sample}

 \vspace{-25pt}
\begin{IEEEbiography}[{\includegraphics[width=0.95in,height=1.2in,clip,keepaspectratio]{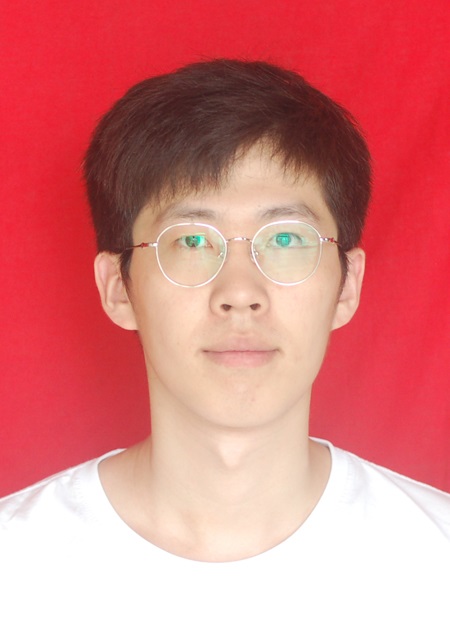}}]{Zhaoyang Cheng}
 received his B.Sc. and B.Ec. degrees from Xi'an Jiaotong University, Xi'an, China, in 2019, and  Ph.D. degree from Academy of Mathematics and Systems Science, Chinese Academy of Sciences, Beijing, China, in 2024. He is currently a postdoctoral researcher with the Division of Information Science and Engineering, School of Electrical Engineering and Computer Science, KTH Royal Institute of Technology, Stockholm, Sweden.  His research interests include game theory, security, and privacy.
\end{IEEEbiography}
 \vspace{-25pt}
\begin{IEEEbiography}[{\includegraphics[width=0.95in,height=1.2in,clip,keepaspectratio]{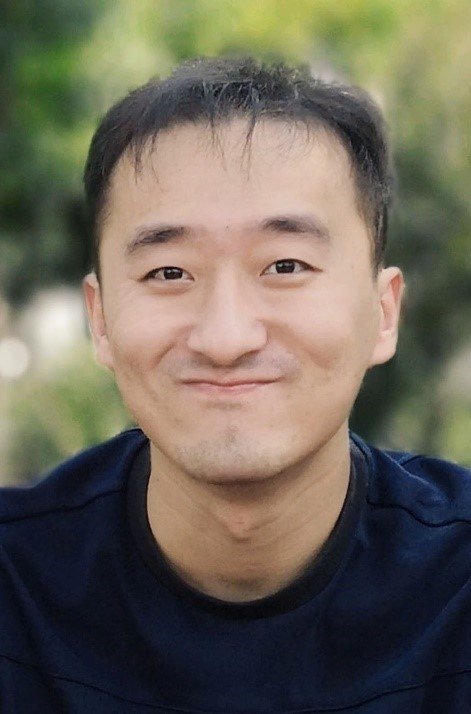}}]{Guanpu Chen} received his B.Sc. degree from University of Science and Technology of China, Hefei, China, in 2017, and Ph.D. degree from Academy of Mathematics and Systems Science, Chinese Academy of Sciences (CAS), Beijing, China, in 2022. He is currently a Professor with the School of Automation, Southeast University, Nanjing, China. He used to be a postdoctoral researcher with the School of Electrical Engineering and Computer Science, KTH Royal Institute of Technology, Stockholm, Sweden. His research interests include multi-agent systems, network games, as well as robustness, resilience, and security in cyber-physical systems. Prof. Chen was the recipient of the President Award of CAS, the Best Paper Award at IEEE ICCA, Guan Zhao-Zhi Award at CCC, the Best Paper Award by Unmanned Systems, and the Best Student Paper Honorable Mention at IEEE CSS TCSP. 
\end{IEEEbiography}
 \vspace{-25pt}
\begin{IEEEbiography}[{\includegraphics[width=0.95in,height=1.2in,clip,keepaspectratio]{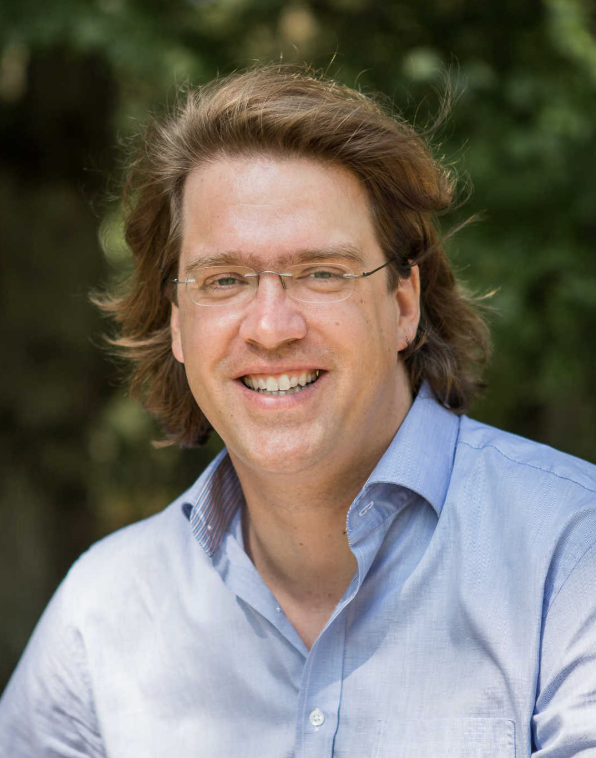}}]{Tobias J. Oechtering}(S’01-M’08-SM’12)  received his Dipl-Ing degree in Electrical Engineering and Information Technology in 2002 from RWTH Aachen University, Germany, his Dr-Ing degree in Electrical Engineering in 2007 from the Technische Universit\"at Berlin, Germany. In 2008 he joined KTH Royal Institute of Technology, Stockholm, Sweden and has been a Professor since 2018. In 2009, he received the ``F\"orderpreis 2009” from the Vodafone Foundation.

Dr. Oechtering is currently Senior Editor of IEEE Transactions on Information Forensic and Security since May 2020 and served previously as Associate Editor for the same journal since June 2016, and IEEE Communications Letters during 2012-2015. He has served on numerous technical program committees for IEEE sponsored conferences, and he was general co-chair for IEEE ITW 2019. His research interests include privacy and physical layer security, statistical learning and signal processing, communication and information theory, as well as communications for networked control. 
\end{IEEEbiography}
 \vspace{-25pt}
\begin{IEEEbiography}[{\includegraphics[width=0.95in,height=1.2in,clip,keepaspectratio]{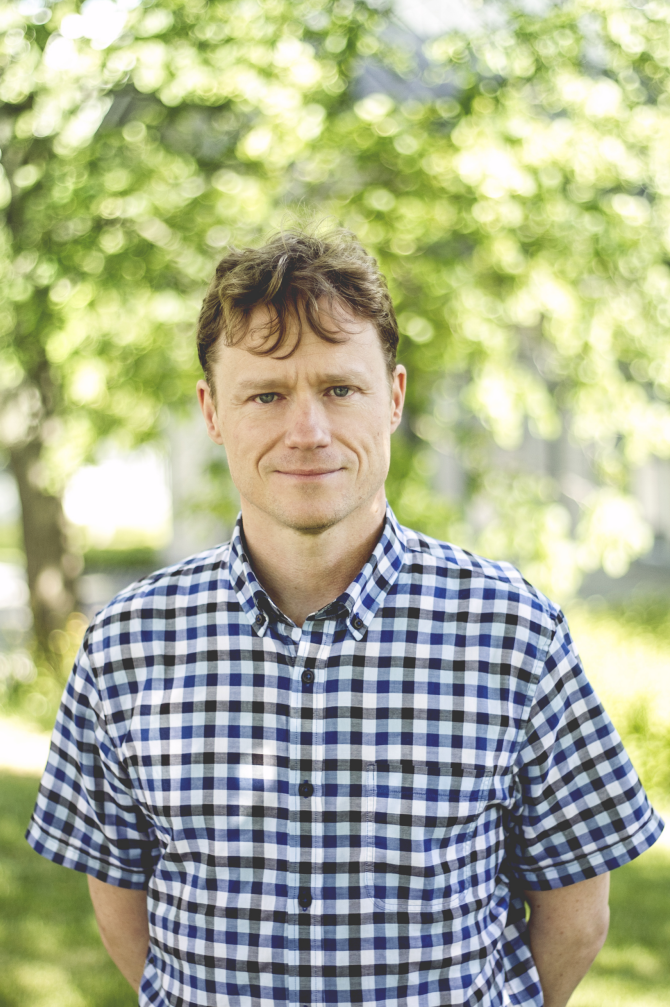}}]{Mikael Skoglund}(S'93-M'97-SM'04-F'19)
received the Ph.D. degree in
1997 from Chalmers University of Technology, Sweden.  In 1997, he
joined the Royal Institute of Technology (KTH), Stockholm, Sweden,
where he was appointed to the Chair in Communication Theory in 2003.
At KTH he heads the Division of Information Science and Engineering.

Dr. Skoglund has worked on problems in source-channel coding, coding
and transmission for wireless communications, Shannon theory,
information-theoretic security, information theory for statistics and
learning, information and control, and signal processing. He has
authored and co-authored more than 200 journal and 450 conference
papers.

Dr. Skoglund is a Fellow of the IEEE. During 2003--08 he was an
associate editor for the IEEE Transactions on Communications.  In the
interval 2008--12 he was on the editorial board for the IEEE
Transactions on Information Theory and starting in the Fall of 2021 
he joined it once again. He has served on numerous technical program
committees for IEEE sponsored conferences, he was general co-chair for
IEEE ITW 2019 and TPC co-chair for IEEE ISIT 2022. He is an elected
member of the IEEE Information Theory Society Board of Governors.
\end{IEEEbiography}

\end{document}